\documentclass[sigconf]{acmart}

\AtBeginDocument{%
  \providecommand\BibTeX{{%
    \normalfont B\kern-0.5em{\scshape i\kern-0.25em b}\kern-0.8em\TeX}}}

\copyrightyear{2024}
\acmYear{2024}
\setcopyright{licensedusgovmixed}\acmConference[CHI '24]{Proceedings of the CHI Conference on Human Factors in Computing Systems}{May 11--16, 2024}{Honolulu, HI, USA}
\acmBooktitle{Proceedings of the CHI Conference on Human Factors in Computing Systems (CHI '24), May 11--16, 2024, Honolulu, HI, USA}
\acmDOI{10.1145/3613904.3642265}
\acmISBN{979-8-4007-0330-0/24/05}

\graphicspath{{images/}}

\usepackage{relsize} % .
\usepackage{soul}

\usepackage{algorithm}
\usepackage{algpseudocode} % .

\usepackage[utf8]{inputenc}
\usepackage{textcomp}
\usepackage{xcolor}

\usepackage{tikz} % .
\newcommand*\circled[1]{\tikz[baseline=(char.base)]{
            \node[shape=circle,draw,inner sep=0.5pt] (char) {#1};}}

\hypersetup{
    colorlinks=true,
    linkcolor=black,
    citecolor=black,
    filecolor=black,
    urlcolor=black,
}

\begin{document}

\title{Color Maker: a Mixed-Initiative Approach to Creating Accessible Color Maps}

%\title{\underline{\href{https://colormaker.org}{\color{black}{ColorMaker.org}}}: a Mixed-Initiative Approach to Creating Accessible Color Maps}

%\author{Anonymous Authors}
%\email{email@email.com}
%\affiliation{%
%  \institution{Institution}
%  \streetaddress{P.O. Box 0000}
%  \city{City}
%  \state{State}
%  \country{USA}
%  \postcode{00000-0000}
%}

\author{Amey Salvi}
\affiliation{%
   \institution{Indiana University Indianapolis}
   \city{Indianapolis}
   \state{IN}
   \country{USA}}
\email{amsalvi@iu.edu}

\author{Kecheng Lu}
\affiliation{%
   \institution{Shandong Univeristy}
   \city{Qingdao}
   \state{Shandong}
   \country{China}}
\email{lukecheng0407@gmail.com}

\author{Michael E. Papka}
\affiliation{%
   \institution{Argonne National Laboratory}
   \city{Lemont}
   \state{IL}
   \country{USA}}
\affiliation{%
   \institution{University of Illinois Chicago}
   \city{Chicago}
   \state{IL}
   \country{USA}}
\email{papka@anl.gov}

\author{Yunhai Wang}
\affiliation{%
   \institution{Renmin University of China}
   \city{Beijing}
   \country{China}}
\email{wang.yh@ruc.edu.cn}

\author{Khairi Reda\textsuperscript{$\ast$}}\thanks{$\ast$ Work performed in part while on sabbatical at Argonne National Laboratory.}
\affiliation{%
   \institution{Indiana University Indianapolis}
   \city{Indianapolis}
   \state{IN}
   \country{USA}}
\email{redak@iu.edu}

\renewcommand{\shortauthors}{Salvi et al.}

% turn off highlgihting by uncommenting the following
\renewcommand\hl[1]{#1}
\begin{abstract}
  Quantitative data is frequently represented using color, yet designing effective color mappings is a challenging task, requiring one to balance perceptual standards with personal color preference. Current design tools either overwhelm novices with complexity or offer limited customization options. We present ColorMaker, a mixed-initiative approach for creating colormaps. ColorMaker combines fluid user interaction with real-time optimization to generate smooth, continuous color ramps. Users specify their loose color preferences while leaving the algorithm to generate precise color sequences, meeting both designer needs and established guidelines. ColorMaker can create  new colormaps, including designs accessible for people with color-vision deficiencies, starting from scratch or with only partial input, thus supporting ideation and iterative refinement. We show that our approach can generate designs with similar or superior perceptual characteristics to standard colormaps. A user study demonstrates how designers of varying skill levels can use this tool to create custom, high-quality colormaps. ColorMaker is available at: \textbf{\href{https://colormaker.org}{\color{blue}colormaker.org}}
\end{abstract}







\begin{teaserfigure}
  \centering
  \includegraphics[width=.85\textwidth]{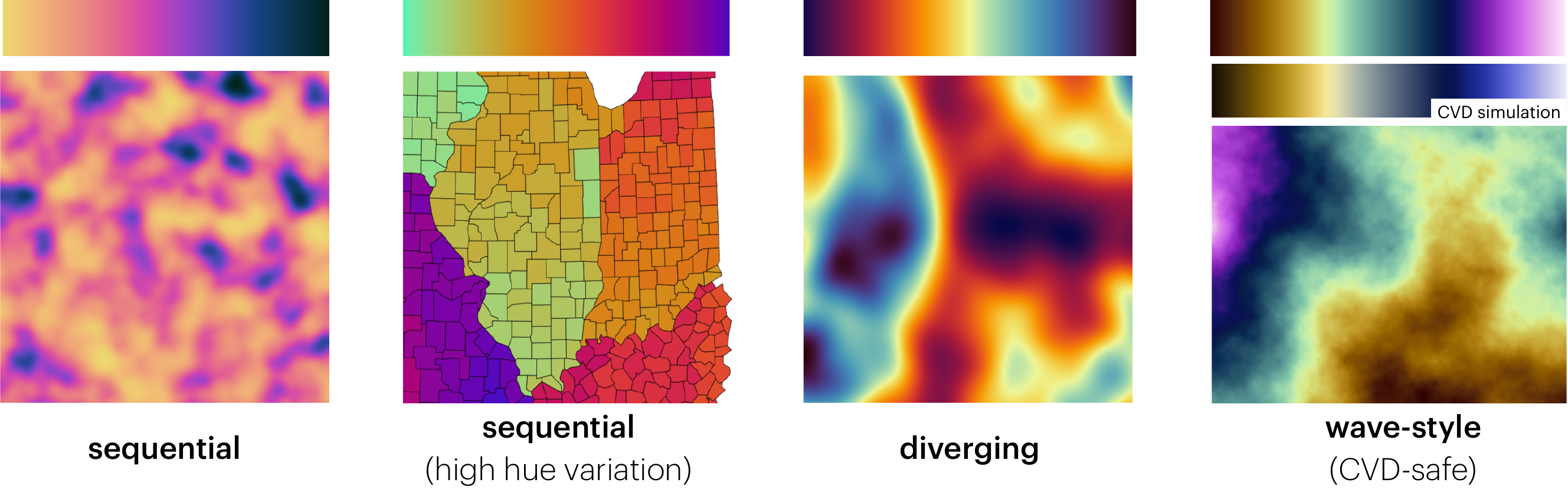}
  \caption{
ColorMaker enables users to create and customize continuous colormaps in a variety of styles, including sequential, diverging, and `wave'-style. Beyond optimizing for fundamental perceptual properties like uniformity, ColorMaker can also ideate and generate designs that are safe for individuals with color vision deficiency (CVD).
  }
  \label{fig:teaser}
  \Description[Four distinct colormaps, paired with a visualization]{
    Four distinct colormaps are displayed, each paired with a scalar field or choropleth below it. The first is a sequential colormap, transitioning from bright yellow to dark blue with shades of pink and purple in between. It is illustrated in a scalar field displayed below it. The second colormap is also sequential but has high hue variation. The color hues included are green, yellow, orange, red, and purple. It is presented in a county-level choropleth, shown below it. The third is a diverging colormap, spanning from dark red to yellow to dark blue. It is displayed using a scalar field. The fourth and final colormap is a wave-style design, which is marked as a CVD-safe colormap. It has the color hues of dark brown to green to purple. Below the colormap, a CVD simulation demonstrates how it appears to those with CVD, with colors transitioning from yellow to grey to blue. Below it is a scalar field, visualizing the colormap.
  }
\end{teaserfigure}

\maketitle

\section{Introduction}

Color is one of the most evocative and readily apparent properties in a visualization. Used effectively, color can help communicate complex quantitative information, such as a hurricane's projected track, the cosmic microwave radiation left over from the Big Bang, or the change in global temperatures over millennia. Despite the importance of this channel to visualization, designing quantitative colormaps is challenging. Visualization designers need to carefully balance various factors, such as perceptual uniformity~\cite{zhou2016survey}, color nameability~\cite{reda2021color}, and personal preferences. Given the complexity of the task, most color advice tools provide a limited set of manually crafted color scales to select from. Novice users are often cautioned against the temptation to use intuitional defaults like the rainbow~\cite{borland2007rainbow}, and instead encouraged to choose from an existing selection of expertly designed colormaps~\cite{harrower2003colorbrewer}. While acceptable in many cases, pre-constructed colormaps lack flexibility, restricting their usage in scenarios that dictate specific design needs. For instance, a visualization designer may desire specific colors at certain scale intervals to meet conventions (e.g., green for plain-level elevations and white for mountain peaks). Meeting these diverse requirements using only predefined colormaps is challenging.

Several techniques exist for generating categorical~\cite{lu2020palettailor,gramazio2017colorgorical,lu2023interactive} or ordered palettes~\cite{petroff2021accessible}, but only a few explicitly support the creation of continuous colormaps. For example, Color Crafter allows users to select a preferred `seed' color~\cite{smart2019color}, which is then fitted to a suitable colormap curve. Although intuitive, curve fitting provides only limited control parameters. Other tools, like CCC~\cite{nardini2019making}, afford direct control over the colormap curve, providing greater design flexibility. However, this approach assumes that users are comfortable specifying their design requirements as a series of control points in a perceptual color space -- a model that may be unintuitive for many, except for power users.

Hence, there is a need for design tools that provide a high level of design customizability without requiring precision specification or expertise in color perception. Moreover, since crafting colormaps for visualization is an iterative process, tools should ideally support users in all design stages, from ideation and evaluation of candidate solutions, to selection and refinement of designs. To address these needs, we introduce \emph{ColorMaker}, a mixed-initiative approach to generating continuous colormaps. ColorMaker formulates the design process as an optimization-based search for a solution that balances user preference, established perceptual guidelines, and visualization accessibility needs. The interface employs familiar drag-and-drop interactions for selecting desired colors. It accommodates approximate or partial design requirements. These specifications are used to build a probabilistic model of user preference, which is then translated into a set of biases and soft constraints onto a simulated annealing optimization. This approach enables interactive generation of colormaps from scratch or with limited input, thus supporting design ideation and generative thinking. ColorMaker also enables the refinement of generated colormaps, automatically operationalizing user edits into additional optimizations to precipitate improved designs. In addition to modeling standard perceptual rules, ColorMaker can also account for color vision deficiency (CVD), generating colormaps that are interpretable by both normal-vision and CVD viewers. 

To evaluate our approach, we compare a large sample of generated colormaps to commonly used designs. We find that ColorMaker generates colormaps with similar or superior perceptual characteristics to established best-practice designs. CVD simulation of generated colormaps also indicates high perceptual discriminability for viewers with CVD. Finally, we conducted a study with visualization designers, demonstrating that ColorMaker is intuitive and provides expressive power for users of all expertise levels. 

\hl{Our \textbf{primary contribution} is a mixed-initiative approach for creating continuous colormaps. Our approach comprises a novel algorithm for colormap generation and optimization, closely integrated with a user-friendly design interface to support iterative colormap development. This integration promotes creative collaboration between the designer and the system, enabling the creation of new, accessible colormaps either from scratch or through partial user specification. Evaluation demonstrates the approach's intuitiveness and ability to produce high-quality colormaps.} We have implemented this technique as an open-source\footnote{The source code is available at: \url{https://osf.io/4ugf6}}, web-accessible tool: \textbf{\color{blue}\url{https://colormaker.org}}.

\section{Related Work}

Various tools have emerged to facilitate the creation and analysis of colormaps. We situate our work in this landscape and discuss related approaches for colormap optimization, modeling, and generation. 

\subsection{Color Design Tools}

For web designers and digital artists, a variety of tools exist, such as Adobe Color CC~\cite{adobe2014farbrad}, Coolors~\cite{coolors2023}, Colormind~\cite{colormind2023}, ColorMagic~\cite{colormagic2023}, and Canva~\cite{canva2023}. These tools offer a range of features, including the automatic generation of harmonious color palettes, sometimes based on an input image. However, these tools cater to general design needs and might not address specific concerns of data visualization, such as perceptual uniformity or discriminability.

Several tools have been designed to scrutinize and evaluate the perceptual properties of colormaps for visualization. For instance, ColorMeasures offers a comprehensive analysis of colormaps, exploring factors such as uniformity, discriminative power, and order in multiple color spaces~\cite{bujack2017good}. However, it primarily focuses on analysis rather than colormap creation. Similarly, cols4all and VizPalette enable users to analyze and compare palettes but lack the capability to generate new designs \cite{tennekes2023cols4all, lumeeks2023vizpalette}.

Certain tools provide data- or task-driven colormap recommendations. For example, PRAVDAColor suggests suitable colormaps based on data type, spatial frequency, and representation goal, aligning its recommendation with the perceptual theory of the time~\cite{rogowitz1996not}. One of the most popular color selection tools is ColorBrewer, which provides high-quality, hand-crafted colormaps that meet design and perceptual standards, but offer limited modification options \cite{brewer1994color}. The CCC tool allows users to create and optimize continuous colormaps~\cite{nardini2019making}. It provides a higher-level specification format that is meant to reduce the number of control points needed for colormap manipulation. Although the CCC tool provides high design flexibility, users are expected to edit and specify control points to realize new designs. This paradigm is arguably challenging for most, save for experienced users. By contrast, ColorMaker separates colormap specification from its user preference model, making it more user-friendly. Similarly, ColorMoves employs a straightforward interface that allows users to build elaborate colormaps by recombining a set of predefined `inserts'~\cite{samsel2018colormoves}. However, unlike ColorMaker, this tool does not provide functionalities to computationally evaluate or optimize designs. Paraview integrates a colormap design dialog into a general-purpose scientific visualization suite, albeit without providing evaluation or optimization options~\cite{ayachit2015paraview}. 

ColorMaker strives to provide greater flexibility than existing color design tools, while still being user-friendly. We take inspiration from recent advancements in visualization authoring environments, including tools like Data Illustrator~\cite{liu2018data}, Charticulator~\cite{ren2018charticulator}, and Lyra~\cite{satyanarayan2014lyra}. Similar to these tools, ColorMaker leverages intuitive drag-and-drop interactions to facilitate creative expression in colormap design. Given the need for precision specification in quantitative colormaps, we combine these user-friendly design features with an algorithm for generative modeling and optimization. Using principles of creative mixed-initiative  interfaces~\cite{horvitz1999principles, deterding2017mixed,heer2019agency}, we enable the user and the system to co-create perceptual colormaps that cater to a wide range of styles and preferences.

\subsection{Computational Colormap Generation and Optimization}

A number of tools employ modeling and optimization-based methods to generate color encodings. ColorCrafter uses clustering techniques to model existing design practices, allowing the latter to be used as the basis for generating new colormaps~\cite{smart2019color}. However, this approach is limited in expressiveness, as users can only select a single `seed' color to customize designs. Similarly, \hl{Wijffelaars et al's approach, although capable of replicating ColorBrewer's designs, allows only for a few control parameters, potentially limiting the set of designs that can be expressed} \cite{wijffelaars2008generating}. Colorgorical employs a model-driven approach to generate categorical color palettes with adjustable scoring functions, providing good customizability but is limited to generating categorical palettes  \cite{gramazio2017colorgorical}. Similarly, Palettailor optimizes categorical palettes using simulated annealing and based on the underlying data characteristics~\cite{lu2020palettailor}. However, Palettailor too is limited to qualitative palettes. Nardini et al. propose a set of algorithms to optimize continuous colormaps~\cite{nardini2021automatic}, which can improve perceptual characteristics, including uniformity, smoothness, and intuitive order. \hl{However, these algorithms are meant to improve a given colormap by making small adjustments to the control points. By contrast, ColorMaker employs a stochastic optimization process to explore a larger swath of the design space, enabling it to ideate new colormaps on its own while also conforming to user preferences when the latter is provided.} \hl{ColorCAT is another tool that enables users to create task-driven colormaps, using design heuristics to match color-mapping strategies with certain classes of tasks}~\cite{mittelstadt2015colorcat}. \hl{However, compared to ColorMaker, which allows users to incorporate arbitrary hue preferences, ColorCAT offers limited customization options.} \hl{Uniquely among existing color design tools, our approach tightly integrates user preferences and interactions with algorithmic optimization, enabling a collaborative human-system colormap design experience. This affords design agency while ensuring adherence to perceptual standards.}

\subsection{Addressing Color Vision Deficiency}
Several color mapping tools notably address CVD, which affects approximately 4\% of the population. \emph{cmasher} offers CVD-friendly scientific colormaps \cite{van2020cmasher}, while VizPalette allows users to simulate the effects of different CVD conditions on palettes~\cite{lumeeks2023vizpalette}. Similarly, Chromaticity~\cite{gramazio2017chromaticity} and cols4all~\cite{tennekes2023cols4all} help evaluate color palette discriminability for CVD.
ColorBrewer includes CVD-safe options among its colormaps~\cite{brewer1996guidelines}. Most of these tools either provide a set of pre-crafted colormaps considered to be CVD-accessible or allow users to evaluate their existing designs. Manual generation and modification are still needed to realize new, CVD-accessible colormaps with these tools. 

Exceptions include IWantHue, which allows for generating CVD-safe categorical palettes~\cite{jacomy2023iwanthue}. Nu{\~n}ez et al's Python module takes an arbitrary colormap and outputs a CVD-safe version~\cite{nunez2018optimizing}. This allows for converting problematic designs (e.g., rainbow) to a more accessible form. The technique works by limiting colormaps to a CVD-safe truncated color space. The resulting designs, however, tend to exhibit dull blue to yellow tones, which may be unappealing for normal-color vision viewers. Instead of truncating the color space, ColorMaker allows selection from the entire displayable gamut but limits the design to color \emph{combinations} that cannot be easily confused. This results in a vibrant colormap that retains discriminability for viewers with CVD. 

\subsection{Perceptual Standards for Colormaps}

A multitude of guidelines have emerged to aid in the design of effective colormaps~\cite{zhou2016survey}. Among these, the principle of smoothness has been consistently advocated to prevent abrupt transitions (or color banding) in the color sequence \cite{borland2007rainbow, moreland2016we}. Perceptual uniformity is another pivotal factor in colormap design~\cite{herman1992color}. A perceptually uniform color sequence ensures that observers can easily distinguish differences in data values as represented by adjacent colors~\cite{bujack2017good}. Perceptual `order' is another commonly cited factor in making intuitive colormaps~\cite{borland2007rainbow}. The latter implies that a viewer should be able to deduce the relative ordering of colors without consulting a legend. ColorMaker explicitly models perceptual uniformity and smoothness. 
\hl{While we do not attempt to optimize for order, the latter is helped by ensuring locally monotonic changes in luminance and by limiting the curvature of the colormap. Although not guaranteeing a perceptually orderable colormap in the absolute sense}~\cite{bujack2018ordering}, \hl{together these two factors appear to yield qualitatively good color ordering.}
Luminance control is often emphasized as a crucial factor for continuous colormaps~\cite{rogowitz2001blair,ware1988color,rogowitz1998data,reda2018graphical}, hence we model the latter as a hard constraint\hl{, in line with the approach used in Matplotlib}~\cite{viridis}. \hl{Within this constraint, we allow the user to dictate the luminance profile (e.g., sequential, diverging, or wavy}~\cite{sciviscolor}).

While crucial to effective colormap design, reconciling these guidelines with user preference is non-trivial. ColorMaker encodes these requirements as a set of scoring functions and soft constraints. It then uses simulated annealing to search for a satisfying solution.
\section{Design Requirements}
\label{sec:design_requirements}

In developing ColorMaker, our aim was to create a tool that supports color mapping by visualization designers of all levels of expertise. We wanted to strike a balance between providing designers with creative control and leveraging computational optimization techniques, so as to ensure that the resulting colormaps adhere to fundamental principles of perceptual color encoding. We discuss the key design requirements we sought to fulfill.

\subsection{Balancing Designer Creativity and Automation}

Effective colormap design often involves reconciling the personal preferences of designers with established perceptual standards, such as ensuring luminance monotonicity and perceptual uniformity. 
\hl{Balancing user color preferences with perceptual standards can be daunting.} Our fundamental hypothesis is that by integrating color selection interfaces with interactive user-driven optimization, we can empower designers with significant control while automating the majority of the colormap generation process. With ColorMaker, designers can specify the hues and colors they wish to incorporate, along with their approximate order. Tedious tasks, such as meticulously positioning control points to achieve a smooth and uniform color gradient, are handled by the algorithm. \hl{Beyond automating laborious aspects of colormap development, the algorithm exhibits creative agency by exploring the space of possible solutions, `filling in' gaps within the user's specification, and responding appropriately to user edits.

This approach borrows from principles of creative, mixed-initiative interfaces}~\cite{deterding2017mixed,yannakakis2014mixed}, \hl{leveraging human creativity and computational techniques to support complex design tasks. Taking inspiration from this philosophy, our aim was to create a system where a human designer and the computer take turns to collaboratively refine the design, constrain the space of possible solutions, and ultimately ensure that resulting colormaps align with user preferences while meeting key perceptual metrics for effective visualization.}

\subsection{Minimal Expertise Required}

ColorMaker is designed to be accessible to non-expert users, enabling them to create customized and effective color encodings. The user interface incorporates familiar user interface elements, such as standard color pickers and drag-and-drop interactions. Instead of demanding precise color specifications, ColorMaker was designed to accept uncertain and imprecise user input. For example, a user can specify an approximate red hue at the desired position in a diverging colormap. The optimization algorithm then calculates the best approximation of this preferred color while ensuring a perceptually sound colormap. Importantly, ColorMaker is forgiving of potentially suboptimal or even problematic input. For instance, if a user prefers two opponent hues in close proximity, which would typically result in a sharp visible boundary, ColorMaker will select a midpoint color instead, or steer the design to incorporate one of the two hues selectively. Similarly, when optimizing for color vision deficiency (CVD), the algorithm self-adjusts problematic color pairs (e.g., greens and reds) by selecting similar hues that can still be distinguished by individuals with CVD.

\subsection{Scaffolding Ideation and Iterative Design}

Creating colormaps is an iterative and creative process. Designers might start with a clear vision or have only a vague concept. To support the exploration of the design space, ColorMaker generates colormap suggestions from the ground up by leveraging its stochastic optimization. Multiple distinctive designs can be synthesized and presented simultaneously to stimulate generative thinking. The tool also accommodates partial input, allowing users to indicate preferences for specific hues at different parts of the color scale. Once an initial solution is generated, ColorMaker provides various interactions for refinement. Users can adjust hue or chroma at arbitrary points in the colormap through lightweight editing functionalities. They can also revise their preferences by repositioning desired colors or adjusting intervals. ColorMaker incorporates these refinements into its user preference model and schedules additional optimization runs to interactively update the design. 

While ColorMaker does not assume specific expertise in color perception, it offers optional features for fine-grained editing and colormap evaluation. Users can view perceptual characteristics of colormaps from within the interface, including changes in luminance, color distance, and smoothness. Advanced users can also view and adjust control points to fine-tune the colormap curve, as needed. These adjustments can be performed through direct drag-and-drop interactions, and are similarly incorporated into subsequent optimization runs.

\subsection{Fostering Accessible Visualization}

A significant portion of visualizations employs color schemes that are difficult to interpret by individuals with color vision deficiency (CVD)~\cite{angerbauer2022accessibility}. This issue arises in part because balancing designer preferences with accessibility needs is challenging. ColorMaker addresses this challenge by considering CVD effects during colormap generation. The optimization process penalizes designs that incorporate potentially confusing color combinations (e.g., isoluminant reds and greens), favoring CVD-safe solutions. ColorMaker automatically handles these constraints, allowing designers to focus on articulating their needs while the algorithm takes care of CVD modeling and optimization.
\section{The Color Maker Interface}

\begin{figure*}[t]
    \centering
    \includegraphics[width=1\textwidth]{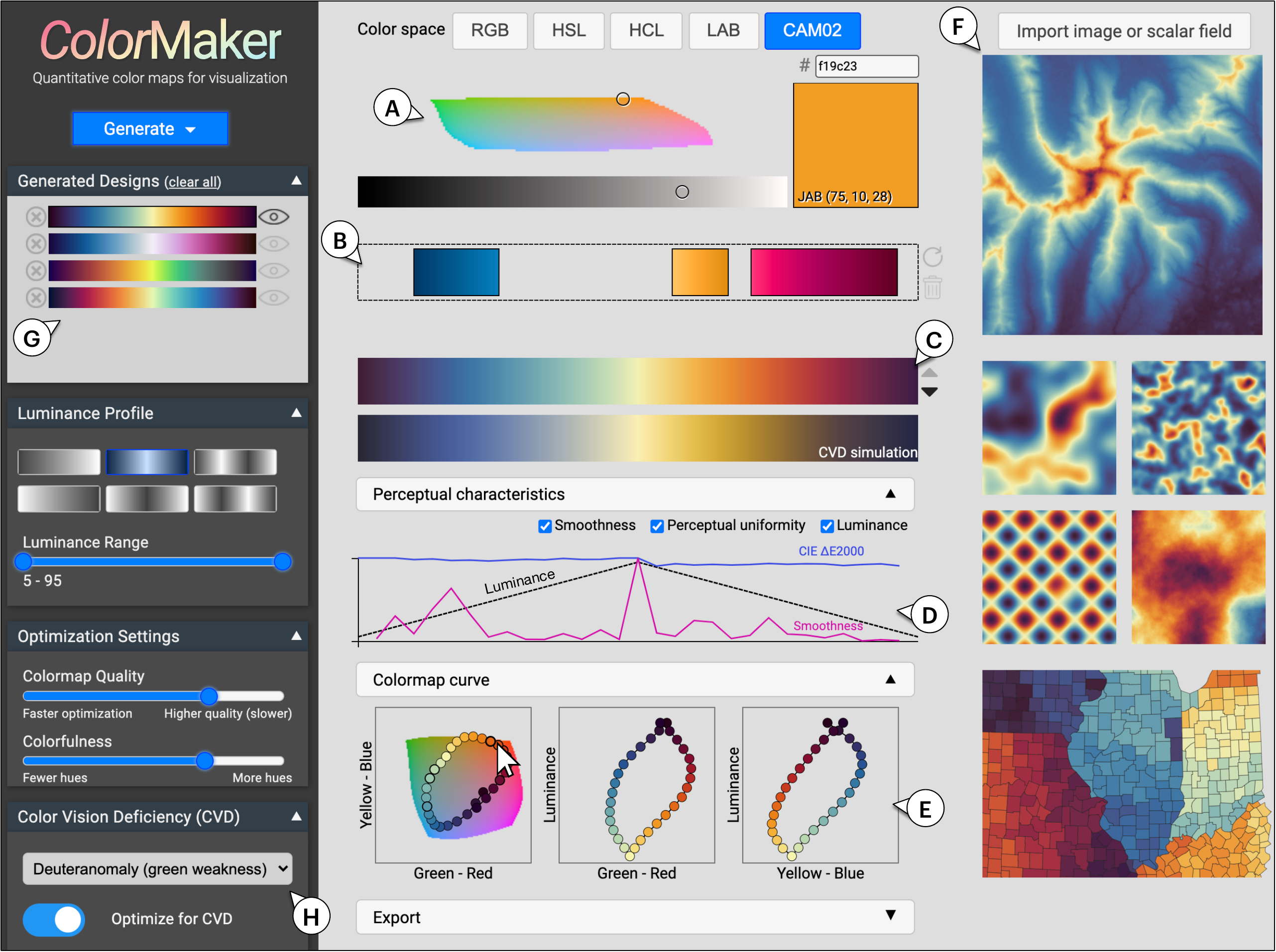}
    \caption{
The ColorMaker interface consists of several key elements: \textbf{(A)} A configurable color picker that supports selection in multiple color spaces. \textbf{(B)} A `preference shelf' that enables users to indicate their preferred colors at their approximate desired positions. \textbf{(C)} The generated colormaps are displayed as continuous scales, accompanied by CVD simulation for accessibility considerations. \textbf{(D)} The user can access information about the underlying perceptual characteristics of the colormap. \textbf{(E)} Users have the option to directly edit the colormap curve if desired, providing fine-grained control. \textbf{(F)} The colormap is applied to a selection of visualizations for quick evaluation. Users can also import their own images or scalar fields for further testing. \textbf{(G)} A sidebar keeps a history of all generated colormaps, allowing the user to revisit and fine-tune previous design variations. \textbf{(H)} Optimization parameters, including the option to simulate CVD effects, can be adjusted to provide additional control over the colormap generation process.}
    \label{fig:ui}
    \Description[Screenshot of ColorMaker's UI]{
    A screenshot of ColorMaker’s UI shows a generated colormap design with its CVD Simulation. The figure also highlights various interactive features and key elements, each identified with an alphabetical indicator. Descriptions of each of these elements marked with their indicators are provided in the caption.
    }
\end{figure*}

ColorMaker is a single, self-contained web application (see Figure~\ref{fig:ui}). We discuss the user interface components and illustrate how they work together with example usage scenarios.

\begin{figure*}[t]
    \centering
    \includegraphics[width=1\textwidth]{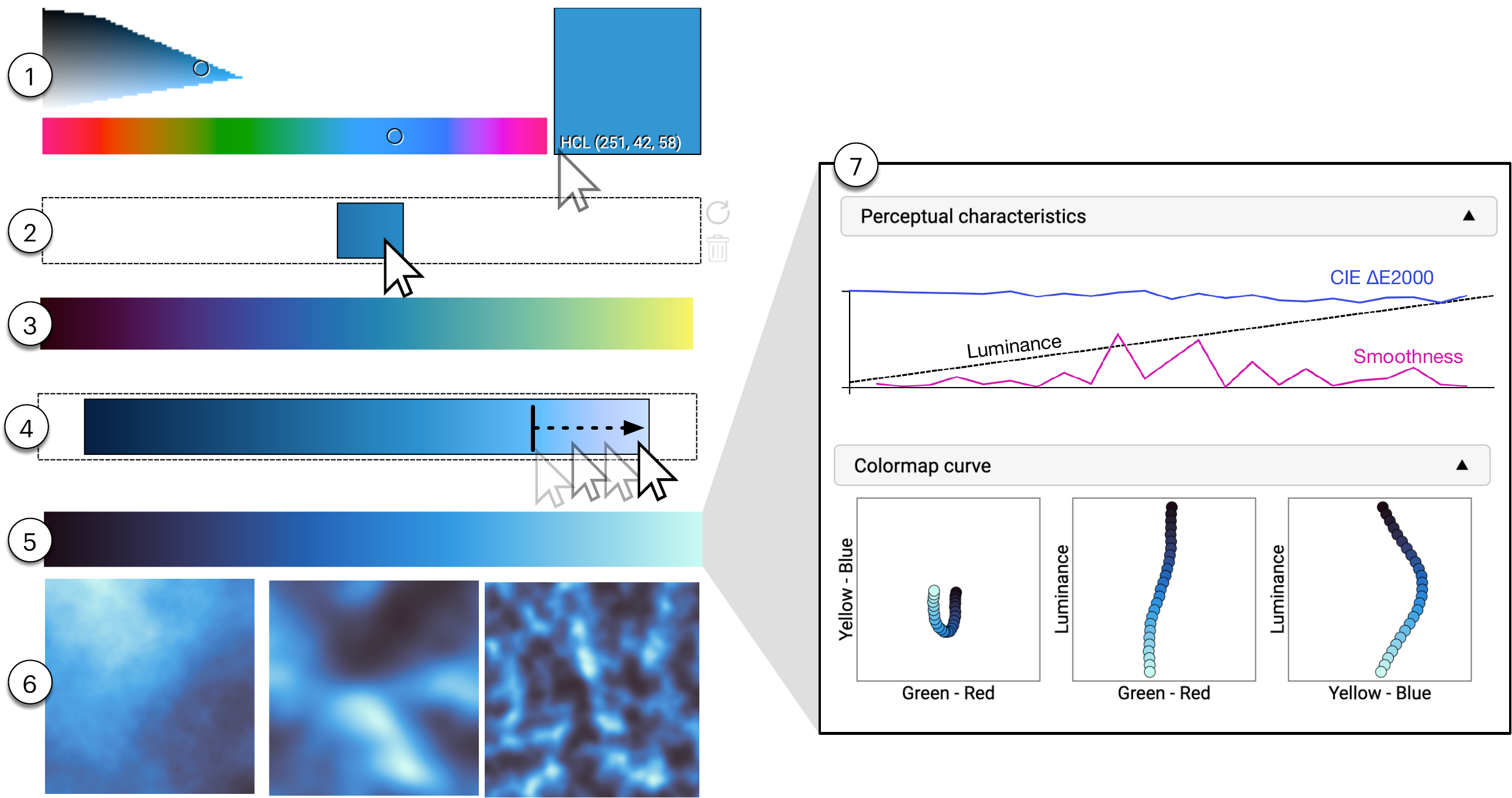}
    \caption{Steps for creating a sequential, single-hue colormap for an ocean bathymetry dataset. A blue color (1) is added to the preference shelf (2), leading to the generation of a multi-hue colormap (3). Stretching of the blue block indicates a preference for a predominantly single-hued design (4), resulting in a second solution that meets the requirement (5). The colormap is checked against example scalar fields (6) and evaluated for its perceptual characteristics (7).}
    \label{fig:usage_scanario1}
    \Description[Steps for creating a sequential, single-hue colormap]{
        The steps for crafting a sequential, single hue-colormap are displayed. The first step shows the color picker with the LCh color space selected, where the user has chosen a blue color with the values (251,42,58). In step 2 we observe the 'preference shelf', where the color selected in step 1 added is positioned centrally in the shelf. Step 3 shows the resulting multi-hue, sequential colormap transitioning from dark to lighter shades. It has a purple hue at the beginning of the colormap, transitioning to a bluish hue near the center, and then shifts to a green and then yellow near the end of the colormap. Step 4 shows the preference shelf again. The mouse pointer indicates the user’s action of stretching and extending the blue color block to encompass nearly the entire shelf. Arrows reveal the direction in which the block was stretched. Step 5 displays a newly generated single-hue blue colormap, transitioning smoothly from darker blue hues to lighter blue hues. In step 6, three scalar fields containing noise and height maps are shown, all visualized using the single-hue blue colormap from step 5. Step 7 highlights the perceptual characteristics of the single-hue blue colormap. This step indicates a linegraph featuring three curves: the monotonically increasing luminance curve ranging from 5 to 95, a uniform curve of CIEDE2000 and a smooth curve with a few minor peaks indicating angular variation. Additionally, colormap curves are presented in the CIE Lab space, with three graphs, A-B, L-A, and L-B, confirming the smoothness of the colormap.
    }
\end{figure*}

\subsection{Incorporating Designer Preference}
\label{sec:pref_shelf}

The ColorMaker interface includes a color picker to facilitate the selection of desired colors (Figure~\ref{fig:ui}-\circled{A}). By default, this picker allows color selection within the \emph{CIE LCh} color space (luminance, chroma, and hue) but also supports several other perceptual (and non-perceptual) color spaces, including CIE Lab, CAM02-UCS, and RGB. \hl{The picker cuts through one color dimension (e.g., $L^*$) via a slider, presenting a 2D slice of the other two dimensions for easy selection.} Once a desired color is chosen, users can simply drag and drop it into a `preference shelf' \circled{B} at the approximate position where they want the color to appear. Each color added to the shelf is represented as a movable color block. Users can add multiple color blocks to convey several preferences simultaneously, such as various hues at different positions within the scale. Furthermore, users can adjust the size of these color blocks to indicate the desired extent within the colormap, as illustrated in Figure~\ref{fig:usage_scanario1}.

The preference shelf \circled{B} serves as the foundation for a user preference model (explained in \S\ref{sec:user_model}). Importantly, ColorMaker intentionally separates this user model from the colormap itself, or more specifically, from the series of control points that define the colormap curve. This separation accommodates fluid preferences, which can be approximate, incomplete, or even capricious. For instance, when adding a preferred color, its position can be chosen arbitrarily. Similarly, the extent of color blocks can be specified flexibly, potentially with multiple colors overlapping. This fluidity is meant to empower designers to think creatively. It is during the optimization process that ColorMaker seeks to find a balance between what the designer wants and what is feasible within perceptual constraints and device limitations (e.g., the color gamut of the display). Whenever user preferences change, such as by moving, adding, or removing preference colors, ColorMaker automatically updates its user model and triggers additional optimization runs to precipitate new designs based on updated preferences. \hl{The resulting colormap is displayed} \circled{C} \hl{along with its perceptual characteristics} \circled{D} \hl{and applied to sample visualizations and test patterns. For additional evaluation, the user can visualize the colormap with a custom dataset} \circled{F}. \hl{The latter can be uploaded as an image or a scalar field.}

\subsection{Refining Generated Designs}
\label{sec:edits}

\begin{figure*}[t]
    \centering
    \includegraphics[width=1\textwidth]{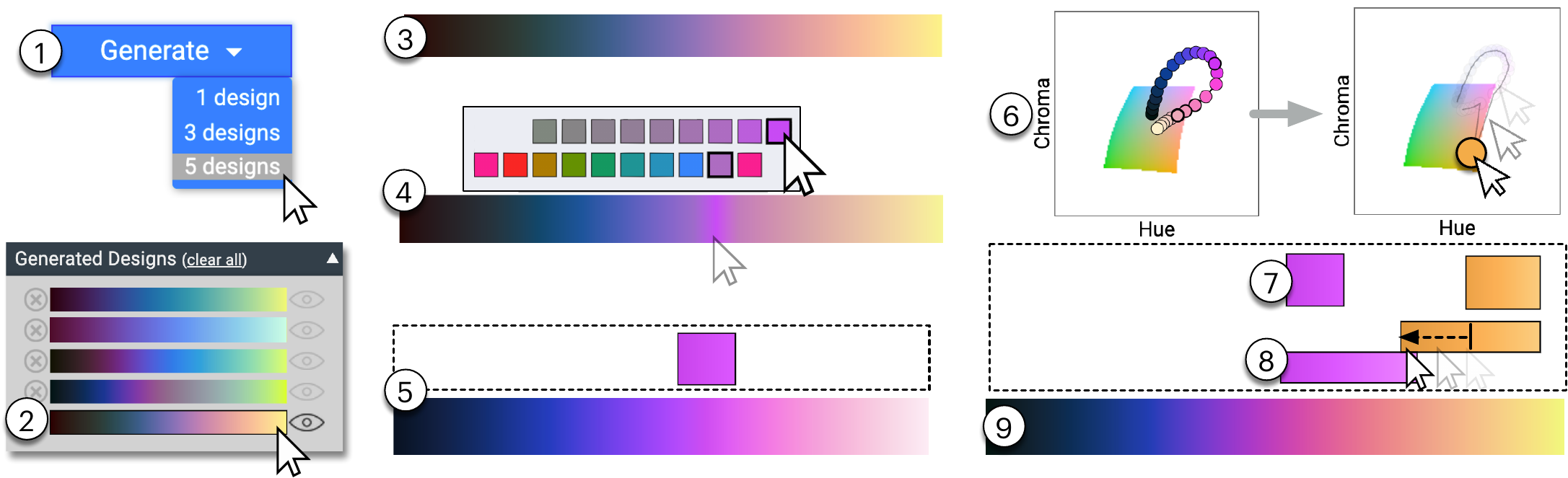}
    \caption{A multi-hue design created by first ideating a random set of solutions (1--2) and selecting one of five generated colormaps to work with (3). The user locally edits the colormap to saturate the purple (4). This adjustment causes the optimization to update the preference model, although it unexpectedly removes the orange-yellow hue previously visible at the end of the scale (5). The user reincorporates that color by editing the curve in the CIE LCh color space (6), and manually repositions the color blocks to indicate a desired blending between purple and orange (7--8), yielding the final colormap (9).}
    \label{fig:usage_scanario2}
    \Description[Steps for creating a multi-hue colormap]{
        The steps for crafting a multi-hue colormap are shown. In Step 1, we observe a 'Generate' button featuring a drop-down arrow, offering options of 1,3, and 5 designs. The user’s cursor highlights the choice of ‘5 designs’. Step 2 shows the 'Generated Designs' tab, displaying the five distinct colormap designs. The fifth design is highlighted as the mouse cursor hovers over it. Step 3 shows an enlarged view of the fifth design from step 2, a multi-hue sequential colormap transitioning from dark blue to purple, further transitioning to a pink and ultimately culminating in an orange-yellow hue. In Step 4, the same colormap from step 3 is displayed with a mouse cursor positioned in the center. Above the cursor, a panel showcases two rows of various colored boxes. The top row shows the variation in chroma (desaturated to saturated) for the pink color highlighted within the colormap. The bottom row features the pink color along with various hues. The cursor highlights a saturated pink color. Step 5 showcases the preference shelf with an added color, representing the pink color selected from the panel in step 4. Below this, a single-hue colormap predominantly in shades of pink is displayed, ranging from dark purples to light pink. In Step 6 we are presented with the colormap curve for Hue vs Chroma, with an overlapping slice of color in the selected luminance. The white color bead from the colormap curve is being dragged towards the orange color. Step 7 reaffirms the changes made in step 6, displaying the preference shelf with the newly added orange color block at the end of the shelf, alongside the existing pink color block positioned near the center. Step 8 shows the preference shelf once more, with the orange and pink color block stretched towards each other, slightly overlapping. In Step 9, we see the newly generated colormap. It is a sequential, multi-hue colormap transitioning from dark blue to purple, vibrant pink, orange and concluding with a yellow color.
    }
\end{figure*}

To refine the generated design, ColorMaker offers two options. First, designers can directly edit the color scale: When users hover over the generated scale, ColorMaker presents a palette of suggested options to adjust the color at the corresponding point, as shown in Figure~\ref{fig:usage_scanario2}-\circled{4}. This palette specifically allows users to modify the chroma and hue of a selected scale point. Users can either increase or decrease the chroma (i.e., saturate or desaturate the color) or choose to rotate the hue. This palette is intentionally limited to a few options that represent localized hue/chroma adjustments for quick refinement. For more precise control, ColorMaker allows users to directly edit the colormap curve by adjusting the position of its control points. For instance, when the CIE Lab space is selected, ColorMaker displays three projections: A-B, $L^*$-A, and $L^*$-B (see Figure~\ref{fig:ui}-\circled{E}). From these projections, users can drag any control point to assign a different color. For example, moving control points in the A-B slice changes the hue and chroma while keeping luminance constant, whereas RGB projections enable adjustments in two of the three color channels while keeping the third constant. To visualize the range of possible color options for a control point, ColorMaker displays a slice of the color space within each projection during user interactions. 

As users refine the colormap using either of the two interactions above, ColorMaker automatically incorporates user edits as additional preferences, integrating them as constraints into the optimization process.

\subsection{Optimization Controls}

In the sidebar, users can select from a variety of luminance profiles, including linear, diverging, and wave-style (seesaw \hl{luminance pattern}~\cite{sciviscolor}), as well as their inverses. Users can also adjust the desired luminance (\emph{L*}) range, which is set to a default of $L^* \in [5, 95]$. The chosen luminance profile is enforced as a strict constraint during the optimization process, ensuring that all generated colormaps adhere to it. Additionally, users can fine-tune other optimization parameters through the sidebar (Figure~\ref{fig:ui}-\circled{H}). This includes a \hl{`colorfulness'} slider that controls the penalty for smoothness; \hl{a higher setting results in a colormap with more curvature, and thus higher hue variation.} Users can also adjust the `colormap quality', which determines the number of iterations in the optimization; higher quality requires more iterations for longer optimization times.

\subsection{Usage Scenarios}
\label{sec:case_studies}

To illustrate how ColorMaker supports colormap ideation and design, we describe three example usage scenarios.

\vspace{1em}\noindent\textbf{Scenario 1 --- Creating a Single-Hue Colormap:} Meet Jane, an Earth scientist working on an oceanography project. She's dealing with a bathymetry dataset, representing ocean bed depth, which naturally progresses from sea level to deeper topographic features. To visualize this dataset, Jane wants a sequential colormap that predominantly uses shades of blue. Opening ColorMaker (Figure~\ref{fig:usage_scanario1}), she selects a blue color from the CIE LCh picker \circled{1}, roughly matching her intended theme. Initially, she places this color somewhat arbitrarily in the middle of the preference shelf \circled{2}. In response, ColorMaker generates an initial colormap \circled{3} that incorporates this particular blue. However, Jane realizes that this initial solution includes additional hues she is not interested in having. To express her preference for a single hue, she stretches the blue color block \circled{4}. This action signals that Jane prefers blue throughout the colormap. ColorMaker reacts by updating its user model and scheduling another optimization. This time, the resulting colormap \circled{5} aligns with Jane's requirements. She checks the example scalar fields \circled{6} provided by ColorMaker, and confirms that this colormap suits her needs. Jane then examines the colormap's properties by opening the `perceptual characteristics' and `colormap curve' panes \circled{7}. She verifies that the colormap indeed has a monotonically increasing luminance profile with even perceptual distances. A glance over the colormap curve in the \emph{CIE Lab} space further confirms that the solution is smooth. Finally, she exports the colormap as an array of control points in the RGB space for convenience.

\begin{figure*}[t]
    \centering
    \includegraphics[width=1\textwidth]{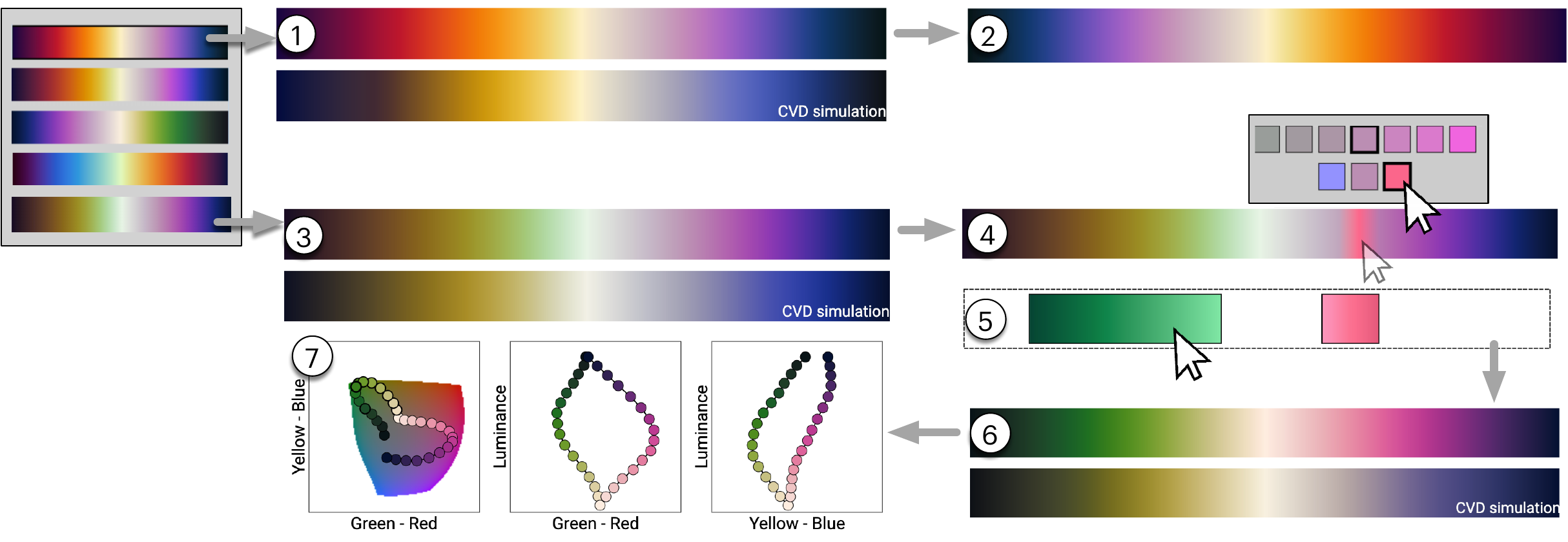}
    \caption{Two diverging colormaps generated for CVD accessibility. Starting from a set of five suggested designs, the user opts for the orange-to-lavender design (1), which is flipped to accommodate desired data semantics (2). As an alternative, the user further selects a second, brown-to-purple colormap (3), refining it to incorporate brighter rose (4) and green (5). Although these colors may be problematic for people with green-red blindness, ColorMaker finds an appropriate solution that balances this preference while maintaining a minimum level of color separability for CVD viewers (6). }
    \label{fig:usage_scanario3}
    \Description[Steps for creating a CVD-safe colormap]{
        The steps for creating a CVD-safe colormap are displayed. In step 1, a panel displays five diverging colormaps, with the first one enlarged and highlighted. This colormap features dark red to red-orange colors on the left arm and purple to dark blue colors on the right, separated by a white hue. Beneath it, a corresponding CVD simulation replicates the colormap with dark brown to yellow in the left arm and light blue to dark blue on the right, separated by a white hue. The text 'CVD Simulation' is written in the lower right-hand corner of the simulation colormap. Step 2 showcases the colormap from step 1 but reversed in the direction, transitioning from blue to white to red. Step 3 provides an enlarged view of the last colormap from the panel, which is also a diverging colormap, ranging from dark brown to green on the left arm, and purple to dark blue on the right, separated by a white hue. The equivalent CVD simulation is shown below, with dark brown to yellow in the left arm and light blue to dark blue on the right, separated by a white hue. The text 'CVD Simulation' appears in the lower right-hand corner of the simulation colormap. In Step 4, the colormap from step 3 is displayed with the mouse cursor hovering over the right arm. Above it, a panel showcases various color boxes. The top row shows the varying chroma of pink colors, from desaturated to saturated. The bottom row presents three colors, purple, light pink, and hot pink, with the mouse cursor highlighting the hot pink. Step 5 shows the preference shelf, featuring hot pink in the right arm, and green in the left arm. The green color is stretched to cover almost the entire left region. Step 6 shows the generated colormap, featuring a diverging colormap transitioning from green to beige to pink and finally to purple. Below it, a CVD simulation illustrates colors ranging from dark brown to yellow to white to dark blue. Step 7 displays the colormap from Step 6 in the form of colormap curves for the CAM02-UCS color space, including curves for A-B, L-A, and L-B. We see a smooth colormap curve.
    }
\end{figure*}

\vspace{1em}\noindent\textbf{Scenario 2 -- Ideating and Refining a Multi-Hue Colormap:} Eduardo is a data scientist who typically uses the default \emph{viridis} colormap in Matplotlib. This time, he is interested in creating a new colormap from scratch. Lacking a clear idea, he utilizes ColorMaker's ideation feature to generate five solutions (Figure~\ref{fig:usage_scanario2}-\circled{1}) without providing any input. ColorMaker responds with five suggested solutions \circled{2} and displays a thumbnail for each. Eduardo previews the designs by hovering over the thumbnails and selects one that intrigues him \circled{3}. While he likes the combination of blue, purple, and orange, he finds the colors somewhat muted. Upon hovering over the scale, ColorMaker suggests alternative chroma and hue values for purple \circled{4}. Eduardo chooses a more saturated color, and ColorMaker responds by adding this selection to the preference shelf \circled{5}, and performing another optimization to fine-tune the design. The new solution predominantly includes the new purple shade, but to Eduardo's surprise, orange has disappeared from the scale. Examining the colormap in the \emph{CIE LCh} space, he notices that the curve has moved up to a more neutral white. To reintroduce orange, he selects a control point near the end of the curve and drags it down \circled{6}, forcing the curve to pass through orange again. This particular orange is automatically added to the preference shelf \circled{7} by ColorMaker. Eduardo further adjusts the preferences shelf \circled{8}, ensuring that the orange and purple blocks overlap slightly for a smooth gradient. Another optimization run accounts for the updated preference model, yielding a colormap \circled{9} that retains the original design features but with more vibrant colors.

\vspace{1em}\noindent\textbf{Scenario 3 -- Creating a   CVD-Friendly Diverging Colormap}: Sue is a climate scientist who is interested in visualizing results from climate simulation. Her model gives projections of temperature fluctuations over the next 100 years relative to a baseline year. To represent temperature differences effectively, she chooses a diverging colormap. However, rather than using the common cool-warm colormap that is typical of climate visualizations~\cite{dixon2022breach}, she is interested in creating a new colormap. To ensure accessibility for colorblind individuals, she activates the `Optimize for CVD' option, specifying deuteranomaly (a common form of color deficiency). She then uses ColorMaker's ideation feature to generate five initial designs (Figure~\ref{fig:usage_scanario3}-\circled{1}). Previewing the suggested designs, Sue selects the orange-to-violet design as a favorite. To match the data semantics (warmer colors for higher temperatures), she flips the colormap horizontally, resulting in a final design \circled{2}.

Sue also explores another suggested design with a dull brown-to-purple profile \circled{3}. To make the colors more vivid, she adjusts purple to a saturated red-rose using ColorMaker's localized suggestion palette \circled{4}. She also adds green \circled{5} to cover the lower arm of the scale. In combination, these two colors can be easily confused by viewers with green-red deficiency. To account for these CVD effects, ColorMaker optimizes by adjusting green to a more neutral beige near the middle of the scale \circled{6}, while retaining a vibrant green in the mid-to-low ranges. Inspecting the colormap curve in the CIE CAM02-UCS space reveals the algorithm has shied away from pure reds, instead incorporating blue-to-purple tones, in effect reducing the likelihood of red-green confusion. While this second solution is arguably less CVD-robust compared to the previous violet-orange design, it demonstrates how ColorMaker can balance competing constraints, finding an acceptable middle-ground solution. 

\vspace{1em}\noindent{In addition to these scenarios, Figure~\ref{fig:teaser} showcases colormaps generated by ColorMaker without specific user preference.}

\section{Method}

Having illustrated the user interface and interactions, we now describe the underlying mechanisms of ColorMaker. ColorMaker uses a custom, simulated annealing algorithm~\cite{aarts1989stochastic} to search the design space. At each iteration, the algorithm randomly perturbs the current solution and evaluates the colormap quality. We score candidate solutions based on their perceptual uniformity, smoothness, and the degree to which they can be considered CVD-accessible. We discuss these scoring functions and the algorithm. We then describe how we incorporate user preferences into the optimization.

\subsection{Assumptions}

A continuous colormap can be described by a function $f: [a, b] \subset \mathbb{R} \mapsto \mathbb{C}$, which maps a scalar $x_i$ to the color $c_i$ in a color space $\mathbb{C}$.  This mapping is defined by an increasing sequence of values $\langle x_1, ..., x_n \rangle \in [a, b]$ and corresponding colors $\langle c_1, ..., c_n\ \rangle \in \mathbb{C}$, with $f(x_i) = c_i$. For simplicity, we assume that the sampling points $x_i$ are equally spaced in the data space (although their associated colors need not be). ColorMaker uses the \emph{CIE Lab} as the underlying color space $\mathbb{C}$. We refer to the sequence of colors C=$\langle c_i \rangle$ as \emph{control points}, and assume that the colormap gradient will pass directly through these colors.  We also assume Euclidean, linear interpolation in the CIE Lab space for mapping values $x$ that lie in between two control points $x_i<x<x_{i+1}$, such that $f(x)=\alpha . \vec{v}_{i+1,i} + \vec{c_{i}}$ where $\vec{v_{j,i}} = \vec{c_{j}} - \vec{c_{i}}$ and $\alpha \in (0,1)$. This in turn allows us to efficiently resample a colormap to \emph{m} control points $C^{\prime}_{(m)} = \langle c^{\prime}_{1}, ..., c^{\prime}_{m}\rangle$. The number of control points $n$, which dictates the degrees of freedom for the optimization, can be varied by ColorMaker depending on the sought luminance profile, with more complex designs (e.g., diverging and wave-style) requiring more control points.

\subsection{Optimization}

We frame colormap generation as an optimization problem to balance three key factors: \textbf{perceptual uniformity}, \textbf{smoothness}, and \textbf{accessibility} for viewers with CVD. These three elements are fundamental to ensuring a quality colormap that can be interpreted by most viewers~\cite{crameri2020misuse,rogowitz1998data}. While other factors, like perceptual order~\cite{borland2007rainbow}, are discussed in the literature, we focus on these three as essential criteria for continuous colormaps. 
%We indirectly promote perceptual order by adopting luminance profiles exhibiting clear direction and by limiting colormap curvature via a smoothness penalty. 
The essence of this optimization is to determine a color sequence $C^*=\langle c_i\rangle$ that minimizes the following objective cost function $E(C)$:

\begin{equation}
C^* = \mathop{arg\min_{C}} \mathbf{E}(C)\text{, where} \nonumber \\ 
\end{equation}

\begin{align}
\mathbf{E}(C) = \omega_u {\mathbf E_{Uniformity}}(C)+ \omega_s {\mathbf E_{Smooth}}(C) + \omega_c {\mathbf E_{CVD}}(C) \label{eq:objective_function}
\end{align}

Here, the three terms represent scoring functions that penalize the colormap for deficiencies in perceptual uniformity (${\mathbf E_{Uniformity}}$), for lack of smoothness (${\mathbf E_{Smoothness}}$), and for the potential of misinterpretation by viewers with CVD (${\mathbf E_{CVD}}$). The $\omega$ terms determine the relative importance of these three factors in the optimization. By default, we set $\omega_u = 0.85$, $\omega_s = 1$, and $\omega_c = 2$. Slightly reducing the weight of perceptual uniformity enhances design diversity while still allowing for uniform designs to emerge (see \S\ref{sec:analysis_perceptual} for results). We assign a higher weight to the CVD factor to increase the likelihood of generating an accessible design. We will further detail each of these three cost functions and discuss the rationale behind their formulation.

\subsection{Perceptual Uniformity}

A perceptually uniform colormap yields a perceived change in color that is proportional to the amount of increase (or decrease) in data values. Thus, a constant change in data should produce a proportional local change in color distance:

\begin{align}
 % \forall i \in \{2, ..., n-1\}: d_{i,i+1}=d_{i,i-1}  \\
 % \text{where}~d_{i,j} = \Delta E^{00}(c_{i}, c_j)
 \forall i \in \{2, ..., n-1\}: d_{i,i+1}=d_{i,i-1}
\end{align}

Where $d_{i,j} = \Delta E^{00}(c_{i}, c_j)$ is the perceived distance between two colors $c_i$ and $c_j$. Here, we use the $CIE ~\Delta E2000$ metric, as it is more accurate than Euclidean Lab distance and can still be computed efficiently. Following Bujack et al.~\cite{bujack2017good}, we quantify the drift from ideal perceptual uniformity by measuring the standard deviation ($\sigma$) of adjacent color distances $d_{i,i+1}$. Because these distances (and their standard deviation) vary considerably across colormap designs (e.g., diverging colormaps exhibit larger deviations than linear), we normalize $\sigma$, dividing by the mean color distance $\bar{d}$. The resulting coefficient of variation ($CV$) represents our normalized perceptual uniformity cost term:

\begin{align}
 \mathbf{E}_{Uniformity}(C) =  {\mathlarger{\frac{\sigma}{\mathlarger{\bar{d}}}}} \label{eq:uniformity}
 = {\mathlarger{\frac{1}{\mathlarger{\bar{d}}}} {\sqrt{ \frac{1}{n-2} {{\sum^{n-1}_{i=1}}}{(d_{i, i+1} -\bar{d})}^{2}}}}  \\
 \text{where}~\mathlarger{\bar{d}} = \frac{1}{n-1} {\sum^{n-1}_{i=1}} d_{i,i+1}
 % {\mathlarger{\sigma\over\mathlarger{\bar{d}}}} \label{eq:uniformity}
 %  = {\mathlarger{1\over\mathlarger{\bar{d}}} {\sqrt{ {1\over{n-2}} {{\sum^{n-1}_{i=1}}}{(d_{i, i+1} -\bar{d})}^{2}}}}  \\
 % \text{where}~\mathlarger{\bar{d}} = {1\over{n-1}} {\sum^{n-1}_{i=1}} d_{i,i+1}
 % \mathbf{E}_{Uniformity}(C) = {\mathlarger{\sigma\over\mathlarger{\bar{d}}}} 
 %= {\mathlarger{n-1\over\mathlarger{\sum^{n-1}_{i=1}} d_{i,i+1}} {\sqrt{ {1\over{n-2}} {\mathlarger{\sum^{n-1}_{i=1}}}{(d_{i, i+1} -\bar{d})}^{2}}}}  
\end{align}

%\begin{gather*}
%    % \textbf{Perceptual Uniformity}(C) = \sum_{i=1}^{n-1} 
%    \textbf{Perceptual Uniformity}(C) = \sqrt{\frac{\sum_{i=1}^{n-1} \left( \Delta E^{00} - \frac{1}{n} \sum_{i=1}^{n} \Delta E^{00} \right)^2}{N}} \\
 %   \text{where}, \; \Delta E^{00} = \Delta E \left(C_{i-1 \over n-1}, C_{i \over n-1}\right)
%\end{gather*}

\subsection{Smoothness}
\label{sec:smoothness}

A smooth colormap transitions seamlessly without abrupt changes in the color gradient. Minimizing color ``banding'' is crucial to prevent viewers from misconstruing false data changes~\cite{borland2007rainbow}. That said, defining a single metric to predict color banding is challenging as there are multiple factors involved, including variations in color names~\cite{reda2020rainbows} and local perceptual distances~\cite{bujack2017good}. However, in a perceptual color space, one indicator of smoothness is the curvature of the colormap curve. A relatively linear colormap tends to incorporate fewer hues while maintaining a nearly constant change in local color distance, resulting in a smoother gradient. Conversely, colormaps with high curvature introduce more hue and chromatic variations, giving rise to visual discontinuities. Additionally, sharp bends in the curve can lead to a reduced perception of color `speed', which manifests as visible bands in the scale.

We therefore assess colormap smoothness by approximating its curvature. An accurate estimate of the latter could be derived from a higher-order representation of the curve, but such computation can be prohibitively costly. Instead, we seek a more efficient approximation by measuring the angle $\theta_i$ between displacement vectors at the control points. Specifically, we approximate colormap curvature as follows:

\begin{align}
 % \text{Smoothness}(C^{\prime}_{(m)}) = { {1\over{m-2}} \sum ({{1-\cos{\theta_i}}\over2})} \label{eq:smoothness}\\ 
 % = {1\over{2(m-2)}}\mathlarger{ \sum_{i=2}^{m-1} ( 1-{{\vec{v}_{i,i-1} \cdot \vec{v}_{i+1,i}}\over{\|{\vec{v}_{i,i-1}\| . \|\vec{v}_{i+1,i}\| } }}) }  \\
 %  \text{where}~\vec{v}_{j,i} = \vec{c}_j - \vec{c}_i
  \text{Curvature}(C) = { {\frac{1}{n-2}} \sum (\frac{1-\cos{\theta_i}}{2}) } \\
  \label{eq:smoothness}
 = \frac{1}{2(n-2)}\mathlarger{ \sum_{i=2}^{n-1} ( 1-\frac{{\vec{v}_{i,i-1} \cdot \vec{v}_{i+1,i}}}{{\|{\vec{v}_{i,i-1}\| . \|\vec{v}_{i+1,i}\| } }}) } \\
 \text{where}~\vec{v}_{j,i} = \vec{c}_j - \vec{c}_i
 %   \text{Curvature}(C) = { {\frac{1}{n-2}} \sum ({{1-\cos{\theta_i}}\over2})} \\
 %  \label{eq:smoothness}
 % = {1\over{2(n-2)}}\mathlarger{ \sum_{i=2}^{n-1} ( 1-{{\vec{v}_{i,i-1} \cdot \vec{v}_{i+1,i}}\over{\|{\vec{v}_{i,i-1}\| . \|\vec{v}_{i+1,i}\| } }}) } \\
 % \text{where}~\vec{v}_{j,i} = \vec{c}_j - \vec{c}_i
\end{align}

Normalization of the cosine term is meant to limit the range of the cost function to $[0, 1]$: 0 indicating a perfectly straight segment (i.e., no penalty), and 1 for a complete reversal of direction (i.e., a $180^\circ$ angle). It is important to note that some curvature is desired so as to allow the colormap to incorporate multiple hues. At times, the designer may wish to limit the number of hues appearing, and hence a higher smoothness penalty may be desired. To meet these needs, we allow the user to partially control the penalty weight, with the full cost function for smoothness defined as: 

\begin{align}
 \mathbf{E}_{Smoothness}(C_{(n)}) = \nonumber \\
  \omega_{s,1} \text{Curvature}(C_{(n)}) + \omega_{s,2} \text{Curvature}(C^{\prime}_{(\lfloor n/2 \rfloor)}) 
 \label{eq:curvature}
\end{align}

The first term computes the penalty at the original resolution of \emph{n} control points with a fixed weight of $\omega_{s,1}=1$. The second term computes the smoothness penalty over a resampled colormap $C^{\prime}$ with $\lfloor\frac{n}{2} \rfloor$ control points. The weight for the latter term $\omega_{s,2} \in [0,1]$ can be controlled by the user. The rationale for the first term is that it is important to ensure a mostly straight line between consecutive colormap segments so as to prevent high-frequency bends. At a lower sampling frequency, the curve can be allowed to deviate from a straight line depending on designer preference (the higher the curvature, the more colorful the design), hence the second, variable-weight term.

\subsection{CVD Accessibility}
\label{sec:cvd}

To optimize for accessibility, we penalize the colormap for incorporating potentially confusable colors. The latter includes certain color pairs that share the same luminance level (e.g., isoluminant green and red tones), which some viewers have difficulty telling apart. To detect problematic colors, we test every pair of equally luminant colors, simulating their appearance to someone with a CVD condition (e.g., deuteranopia). We employ the model proposed by Machado et al. to transform colors from normal-vision coordinates to their simulated  CVD appearance~\cite{machado2009physiologically}. We then measure the distance between the CVD-transformed pairs, and exact a penalty when that distance is lower than a threshold $\gamma({K})$. Specifically:

\begin{figure}[t]
    \centering
    \includegraphics[width=1\linewidth]{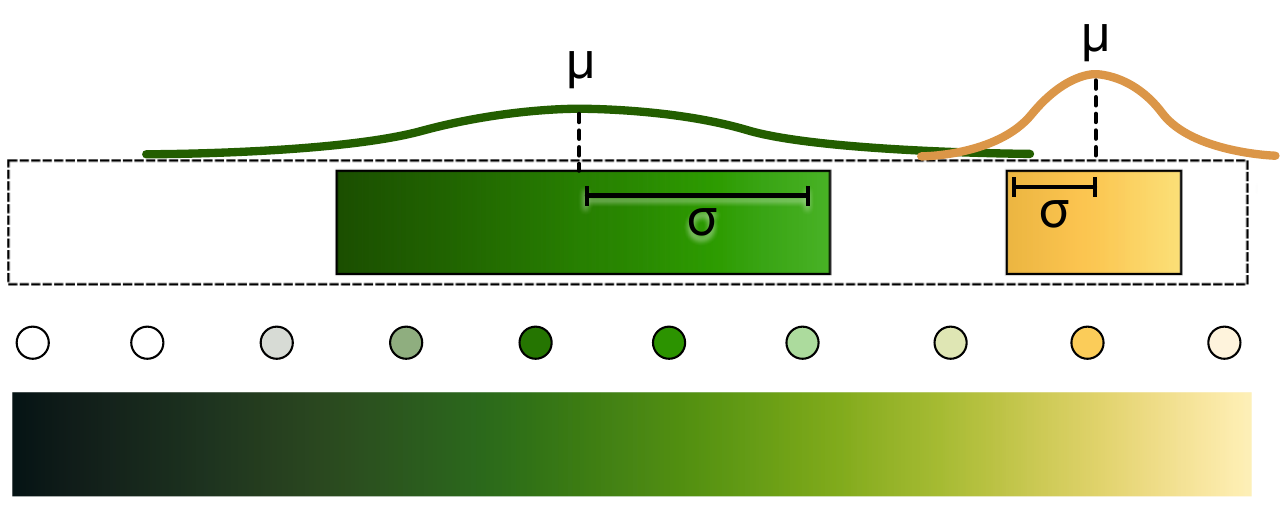}
    \caption{Illustration of how ColorMaker incorporates chromatic bias towards preference colors (green and yellow, in this example). Each color block exerts a certain amount of bias on every control point (circles). The magnitude of the bias follows a normal distribution centered at the middle of the block. In this example, green exerts a more diffused bias impacting several control points, whereas the bias to yellow is focused on the second control point from the right. The cumulative, averaged bias vectors are blended with random offset vectors during the optimization.}
    \label{fig:user_bias}
    \Description[Illustration of chromatic bias towards preference colors in ColorMaker]{
        The preference shelf is shown featuring two color blocks: a green color block positioned near the center, which has been stretched on either sides, and another yellow color block placed at the right end of the colormap. Connecting from the center of these blocks to the ends, a line indicates the sigma value. For both color blocks, we see a normal distribution with the value mu. The distribution for the yellow color is taller than that of the green color block, while the green distribution has a wider spread as compared to the yellow. Below the shelf, a series of 10 circles are arranged equidistant from each other. The circle directly below the color block shares the same color as the block itself, but the colors gradually fade as the circles move away from the center of the block. Below these circles is a sequential colormap. It transitions from a dark green hue to a lighter green, eventually concluding in a light yellow hue.
    }
\end{figure}

\begin{align}
 \label{eq:cvd_accessibility}
 \mathbf{E}_{CVD}(C) = \frac{1}{{\lvert\Psi\rvert}} \mathlarger{\sum_{\substack{\langle i,j \rangle \in \Psi}} \phi(\hat{c_i}, \hat{c_j})} \\
 \text{where } \mathlarger{\hat{c_i}} = T_{CVD} . \mathlarger{c_i} \\
 ~\phi(\hat{c_i}, \hat{c_j}) = 
 \begin{cases}
    0,&  \text{if } \mathlarger{\Delta E(\hat{c_i}, \hat{c_j}) >= K_{ij}} \\
    \mathlarger{1- \frac{\Delta E(\hat{c_i}, \hat{c_j})}{K_{ij}}},&  \text{if } \mathlarger{\Delta E(\hat{c_i}, \hat{c_j}) < K_{ij}}
\end{cases} \\
 K_{ij} = \gamma(K,i,j) = K.\frac{e^{|i-j|/(n-1)}-1}{e-1}
\end{align}

\begin{algorithm}
\caption{Simulated annealing in ColorMaker}\label{alg:cap}
\begin{algorithmic}[1]

\Function{computeUserPrefBias}{$index$, $C_j$}
    \State Set $P$ to the array of preference colors (\S\ref{sec:user_model})
    \State $\vec{u} \gets [0, 0, 0 ]$
    \For {$i \gets 1 \text{ to } length(P)$} \do \\
        \State $\mu \gets P_i.center$; $\sigma \gets \frac{1}{2}  P_i.size$
        \State $\vec{b} \gets [0, P_i.A - C_j.A, P_i.B - C_j.B]$
        \State $\vec{u} \gets \vec{u} + \Call{dnorm}{index, \mu, \sigma} \vec{b}$
    \EndFor
    \State \Return $normalize(\vec{u})$
\EndFunction
    \\
    \State Set $n$ to the number of control points
    \State Set $C$ to an array of $n$ CIE Lab colors
    \State Set initial $T_{init}$ and final $T_{end}$ temperatures, $alpha$ as the cooling rate, and $iterCount$ as the number of iterations \\
    \For {$i \gets 1 \text{ to } n$} \do \\
         \Comment{Initialize $C_i$ to a random Lab color}
        \State $C_i.L^* \gets L^*$ of desired luminance profile at index $i$
        \Repeat 
            \State Set $C_i.A$ and $C_i.B$ to random chromatic values
            %\State $C_i.A \gets random(-150, 150)$
            %\State $C_i.B \gets random(-150, 150)$
        \Until {$C_i$ is within display gamut}
    \EndFor
    \\
    \State $T \gets T_{init}$     %\Comment{Begin optimization loop}

% \State $final \: temperature \gets T_{fin}$
% \State $cooling \: rate \gets \alpha$
% \State $iteration \: count \gets count$
    \State $C_{best} \gets C; E_{best} = E(C)$ \Comment{Set current best solution}
    \While{$(T \geq T_{end})$}
        \For {$i \gets 1 \text{ to } iterCount$} \do \\
            %\State $C_1$ \gets $C_0$
            \State $j \gets \lfloor random(1,n+1) \rfloor$ \Comment{Select an index to perturb}
            \State $\vec{u_j} \gets \Call{computeUserPrefBias}{j, C_j}$ %\Comment{Compute a bias to user preferences at index $j$}
            \Repeat 
                \State $\vec{r} \gets normalize(~[0, random(0,1), random(0,1)]$~)                 \State $\vec{o} \gets normalize(~0.4 \vec{r} + 0.6\vec{u_j}~)$                 %\Comment{Blend random and preference bias vectors}
                \State offset $C_j$ by a small amount in the direction of $\vec{o}$
            \Until {$C_j$ is within display gamut}
            
            %\State \vec{r} \gets 
            %\State randomly perturb one of control points in $C_1$
            \State $E_{new} \gets E(C)$
            \Comment{Cost of the new solution}
            \State $\Delta \gets E_{new} - E_{best}$
            \If{$\Delta \leq 0 \text{\textbf{ OR} } random(0, 1) < 1/(1+e^{\frac{\Delta}{T}})$} %\Comment{If new solution is objectively better OR a under a random draw it should be adopted anyway}
                \State $C_{best} \gets C$
                \Comment{Adopt new as current best}
                \State $E_{best} \gets E_{new}$ 
            \Else
                \State $C \gets C_{best}$
                \Comment{Revert to last best solution}

          \EndIf
        \EndFor
        \State $T \gets T * \alpha$
    \EndWhile
\end{algorithmic}
\end{algorithm}

$\Psi$ is a set of control point pairs that lie on the same $L^*$ slice in the {CIE Lab} space. We obtain a CVD simulated color $\hat{c_i}$ by transforming $c_i$ with a matrix $T_{CVD}$. ColorMaker allows the user to select which CVD condition to simulate and optimize for, with deuteranomaly (the most common) being the default. 
% ColorMaker allows the user to select which CVD condition to simulate and optimize for, with deuteranomaly (the most common) being the default. 
The constant $K$ dictates the minimum distance threshold in CIE Lab units. By default, we use a value of $K=70$ which we found to be a sweet spot. The cost function applies a penalty if the distance between the simulated color pairs, $\hat{c_i}$ and  $\hat{c_j}$, is less than ${K_{ij}}=\gamma(K, i, j)$. The attenuation function $\gamma$ serves to reduce the minimum required threshold $K$ when the two colors $i$ and $j$ are in close proximity within the sequence. This attenuation is important so as not to penalize neighboring colors that should, by necessity, be perceptually similar. \hl{The above cost function ensures continuous and automatic steering of colormaps towards accessible color combinations.} That said, CVD modeling is optional and can be disabled by setting $T_{CVD}$ to an identity matrix $I_3$.
% CVD modeling is optional in ColorMaker and can be disabled by setting $T_{CVD}$ to an identity matrix $I_3$. 
In this case, we still retain the cost function to ensure sufficient distance between the control points, effectively preventing the solution from looping and `reusing' the same colors in different parts of the scale (e.g., using the same blue twice in the two arms of a diverging colormap). 

\hl{One limitation in the current implementation is that it only considers one CVD condition at a time, although the user can specify which condition to model. It is possible to extend Equation}~\ref{eq:cvd_accessibility} \hl{to multiple conditions by summing up the penalties from three CVD matrices, for instance. Doing so, however, severely limits the color space, leading to dull designs. Our approach is to instead model the most prevalent and severe CVD conditions by default while still allowing optimization for less common conditions when needed.}

%\begin{gather*}
 %   \textbf{Smoothness}(C) = \sum_{i=1}^{n-1} \cos\left(\frac{\math{A} \cdot \math{B}}{\|\math{A}\| \|\math{B}\|}\right) \\
    % (C_{i-1 \over n-1}, C_{i \over n-1}) \\
 %   \text{where}, \;
 %   \text{A} = \math{x}_1 + \math{y}_1 + \math{z}_1, \; \text{B} = \math{x}_2 + \math{y}_2 + \math{z}_2, \\
 %   \|\text{A}\| = \sqrt{\math{x}_1^2 + \math{y}_1^2 + \math{z}_1^2}, \; \|\text{B}\| = \sqrt{\math{x}_2^2 + \math{y}_2^2 + \math{z}_2^2}, \\
  %  \math{A} \cdot \math{B} = \text{Dot product of A and B}

%\end{gather*}

\begin{figure*}[t]
    \centering
    \includegraphics[width=.85\textwidth]{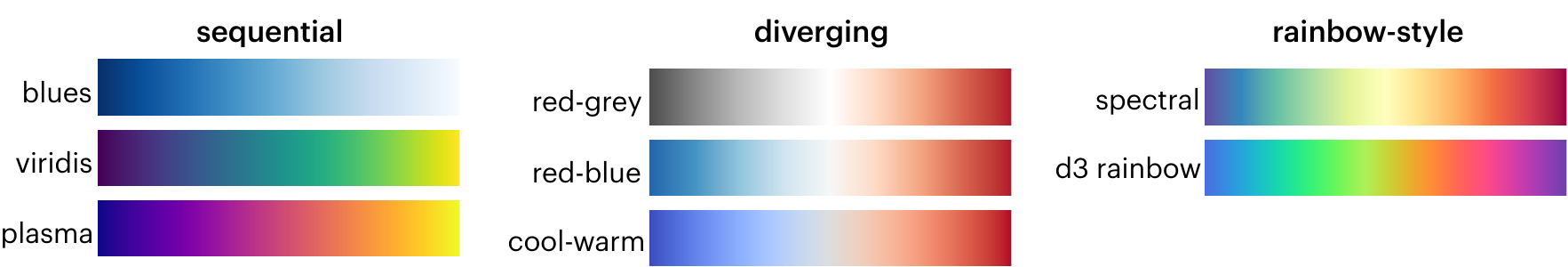}
    \caption{We selected eight commonly used colormaps as benchmarks to compare against. This selection covers three typical design styles: sequential, diverging, and rainbow.}
    \label{fig:benchmarks}
    \Description[Eight colormaps used as benchmarks]{
        We observe eight different colormaps, divided into three columns. The first column is sequential which has three colormaps: ‘Blues’, a sequential single-hue blue colormap; ‘Viridis’, a sequential multi-hue colormap with hues transitioning from purple to blue to green to yellow. And ‘Plasma’, a sequential multi-hue colormap with hues transitioning from dark blue to purple, pink, orange and ultimately yellow. The second column is diverging, which also has 3 colormaps: ‘Red-Grey’ which is a diverging colormap which transitions from dark greys to white to dark red. ‘Red-Blue’, with blues in the left arm and reds in the right arm, separated by a white color, and ‘Cool-Warm’, blues in the left arm, reds in the right arm, separated by a light grayish color, which is much smoother than the other two diverging colormaps. The third column is rainbow-style, which has 2 colormaps. The first is ‘Spectral’, a rainbow colormap that transitions through a spectrum of hues, including blues, greens, and reds. The second is ‘d3-rainbow’, which offers a more vibrant and colorful rendition compared to ‘Spectral’. ‘Spectral’ colormap maintains a more ordered approach to its hue transitions.
    }
\end{figure*}

\subsection{Algorithm}

We employ a custom, simulated annealing algorithm to optimize the objective cost function. The algorithm starts at a random solution with all control points initialized to random colors chosen from the displayable {CIE Lab} gamut. Specifically, we set the luminance of the control points to the requested luminance profile (e.g., monotonically increasing or diverging), while randomly initializing the A and B (i.e., chromatic) components of the color. The algorithm begins at a high initial `temperature', which is then progressively cooled. At every iteration, the algorithm perturbs the current solution by offsetting a randomly selected control point by a small amount, in the direction of a random vector $\vec{r}$. We only offset the A and B components of the color, leaving the $L^*$ channel unchanged. The resulting colormap $C$ is evaluated using the cost function in Equation~\ref{eq:objective_function}. If the new solution yields a lower cost, it is adopted. If the new solution is worse, it may still be adopted with a probability:

\begin{align}
{p=1/{(1+e^{ \mathlarger{ \frac{E(C)-E(C_{best})}{T}} })}}
\label{eq:p}
\end{align}
Where $E(C)-E(C_{best})$ is the difference in cost between the new and the current best solution, and $T$ is the temperature. The scheduling function above, which is commonly used in simulated annealing problems~\cite{henderson2003theory}, ensures a certain probability of accepting a \emph{worse} solution. This behavior allows the algorithm to explore a wider solution space, especially in the earlier, high-temperature cycles of the optimization. Based on prior work~\cite{lu2020palettailor,lu2023interactive}, we choose an initial temperature value of ${T_{init}} = 1$ and final temperature $T_{end}=0.0001$, with a cooling coefficient $\alpha = 0.925$. We set the number of iterations at each temperature level to $iterCount =$ 5,500. This number can be increased or decreased by a slider in the user interface, allowing for either faster or slower (and thus higher quality) optimization, as needed. The pseudo-code is shown in Algorithm~\ref{alg:cap}.

\subsection{Operationalizing User Edits and Preferences}
\label{sec:user_model}

The optimization iteratively perturbs the colormap, one color at a time until it converges to a good solution. The perturbation is, by design, stochastic. However, to incorporate user preferences, we can bias the algorithm to favor colors specified by the user. This is done by blending the random vector $\vec{r}$ with a bias vector $\vec{u}$ pointing in the direction of color(s) included in the preference shelf (see \S\ref{sec:pref_shelf}). Specifically, for a control point $c_j$ selected for perturbation, the vector $\vec{u_j}$ will point toward the color (or average of multiple colors) indicated by the user as preference. The magnitude of bias is dictated by the extent of a color block $2\sigma$, and the distance between the control point and the midpoint of the color block $\mu$: the closer $j$ is to the center, the stronger the bias (see Figure~\ref{fig:user_bias}). Bias magnitude is also modulated by the extent of the preference: a narrower color block exerts a more focused bias concentrated at its center, whereas a wider block conveys a more diffused bias that impacts more control points although to a lesser extent. The overall bias vector $\vec{u_j}$ is an accumulation of weighted biases, which are then blended with $\vec{r}$ at a 60-40\% ratio, respectively (line \#31 in Algorithm~\ref{alg:cap}). In effect, we allow the algorithm to favor colors indicated by the designer while still leaving a sufficient level of randomness in the optimization.

The above model of \emph{designer preference as bias} enables ColorMaker to operationalize initial preferences and post-optimization refinement in a uniform manner. Specifically, edits, such as adjusting the chroma or hue (see \S\ref{sec:edits}), are incorporated into the preference shelf and factored into the optimization according to the model above. One difference is that, when re-executing the optimization as a result of edits, we use the previous solution as a starting point, in lieu of a random initial solution. Additionally, we also initialize the algorithm to start at a lower temperature. This enables us to maintain the global structure of the last solution while still responding to local edits. The lower starting temperature also allows for a faster, more interactive optimization. That said, the user can choose to run a full optimization cycle by clicking a `re-run' button.

\section{Evaluation}

We evaluate ColorMaker in two ways: First, by analyzing the perceptual characteristics of a representative sample of generated design, and second, through a user study conducted with visualization designers and researchers.

\begin{figure*}[t]
    \centering
    \includegraphics[width=1\textwidth]{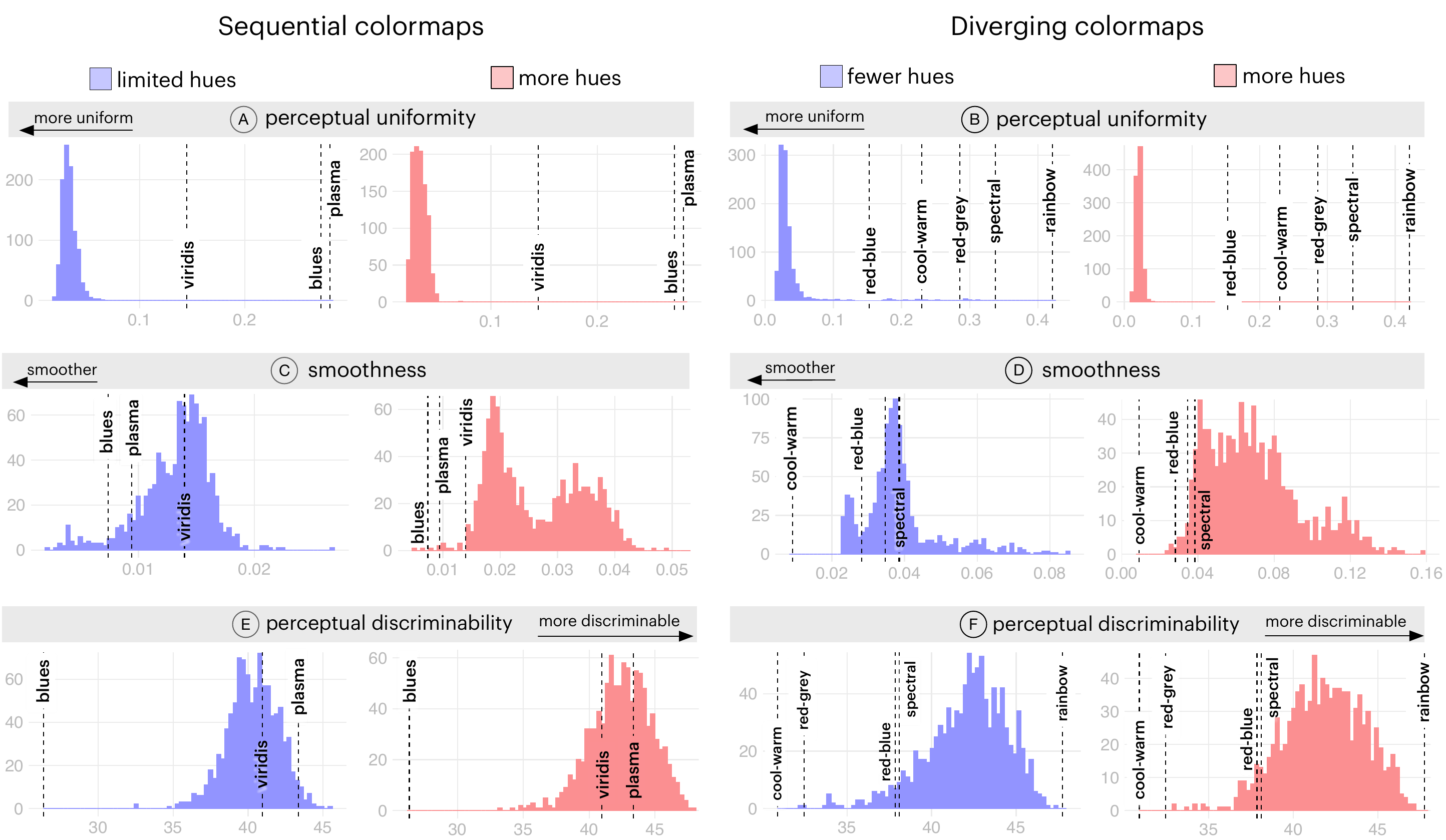}
    \caption{Comparison of perceptual characteristics for generated designs vs. comparable benchmark colormaps. Histograms represent a sample of 1,000 colormaps generated either as sequential (left) or diverging (right). Blue and red represent optimization for either limited or unconstrained hue variation.}
    \label{fig:perceptual_quant}
    \Description [Comparison of perceptual characteristics for generated designs vs. benchmark colormaps]{
        A two-column format is presented, with each column comprising three rows, one for each perceptual characteristic. Within each row, there are two histograms, one representing limited hues (depicted in blue) and the other representing more hues (depicted in red). In the first column, which focuses on sequential colormaps, the first row examines perceptual uniformity. In the row, the blue histogram is narrower and positioned towards the left side, indicating limited hue variability. The benchmark colormap ‘Viridis’, indicated by a dotted vertical line, is located towards the center-right, while ‘Blues’ and ‘Plasma’ are positioned towards the extreme right. As we move to the left in the graph, uniformity increases. A similar pattern is observed for the red histogram. For the second row, representing smoothness, displays a wider histogram, slightly left-skewed. ‘Blues’ and ‘Plasma’ are to the left of the histogram's center, while ‘Viridis’ is at the center of the histogram. The red histogram is also widely spread but exhibits two peaks and a right skew. All three benchmark colormaps are situated to the left of the leftmost taller peak. Smoothness increases as we move left in the graph. In the final row, which assesses perceptual discriminability, both histograms are slightly left-skewed. The ‘Blues’ colormap indicator is at the extreme left, whereas the ‘Viridis’ and ‘Plasma’ indicators are located towards the center near the peak. Discriminability increases as we move towards the right in the graph. In the second column, the first row is similar to the first column, with the histogram positioned at the extreme left and the benchmark colormap indicators on the right side of the histogram. The second row in the second column mirrors the first column’s pattern, with the benchmark colormaps positioned near the peak for both the blue and red histograms. In terms of discriminability, the third row in the second column showcases the benchmark indicators to the right of the histogram's peak, for both blue and red histograms. The only exception is the ‘Rainbow’, which is situated at the extreme right, indicating higher discriminability.
    }
\end{figure*}

\subsection{Analysis of Perceptual Characteristics}
\label{sec:analysis_perceptual}

In line with previous work~\cite{wijffelaars2008generating,smart2019color}, we evaluate the quality of ColorMaker's algorithm by comparing its output to popular colormaps. We specifically test if our algorithm can generate colormaps that are perceptually similar to established, best-practice designs. To do this, we generate a large sample of colormaps for each of our test cases. We vary the generation parameters, testing sequential and diverging profiles, low vs. unconstrained hue variation, and CVD-optimization vs. non-optimized generation. We evaluate three characteristics: perceptual uniformity (computed according to Equation~\ref{eq:uniformity}), smoothness (Equation~\ref{eq:smoothness}), and perceptual discriminability. The latter is a measure of how distinctive individual colors are relative to all other colors in the scale~\cite{bujack2017good}, computed using the following equation:

\begin{align}
\text{{Perceptual discriminability}}(C) =\frac{2}{n(n - 1)} \mathlarger{\sum_{\substack{i<j\leq n}}\Delta E^{00}(c_i, c_j)}
\label{eq:discriminability}
\end{align}

\hl{This metric measures whether the scale contains colors that are potentially confusable, making it useful to gauge colormap accessibility through simulation, in addition to normal-color vision discriminability.} 

We selected eight benchmark colormaps (see Figure~\ref{fig:benchmarks}) from a range of reputable sources, including Matplotlib~\cite{viridis}, ColorBrewer~\cite{harrower2003colorbrewer}, D3~\cite{bostock2011d3}, and one colormap due to Moreland~\cite{moreland2009diverging}. This selection represents designs often considered by the visualization community to embody good perceptual characteristics. We added D3's rainbow as an upper limit in terms of colorfulness. \hl{The latter exhibits a notably smoother profile than traditional rainbows.}

\hl{We investigate ColorMaker's generation capability focusing on two designs: sequential and diverging colormaps, both commonly used in quantitative visualizations. We also explore high and low colorfulness settings, which directly affect the smoothness penalties detailed in} \S\ref{sec:smoothness}. \hl{To illustrate the impact of this user-controlled parameter, we show two distributions: one with low colorfulness for limited hue variation ($w_{s,2}=0.9$) and another with more varied colors ($w_{s,2}=0.25$, ColorMaker's default value). Additionally, we introduce slight randomness in luminance range $[L^*_0, L^*_1]$ to promote design diversity.} The range was randomized from $L^*_0 = \text{random}(5,15)$ to $L^*_1 = \text{random}(85,95)$, similar to the luminance profiles found in the benchmarks. 

\begin{figure*}[t]
    \centering
    \includegraphics[width=.85\textwidth]{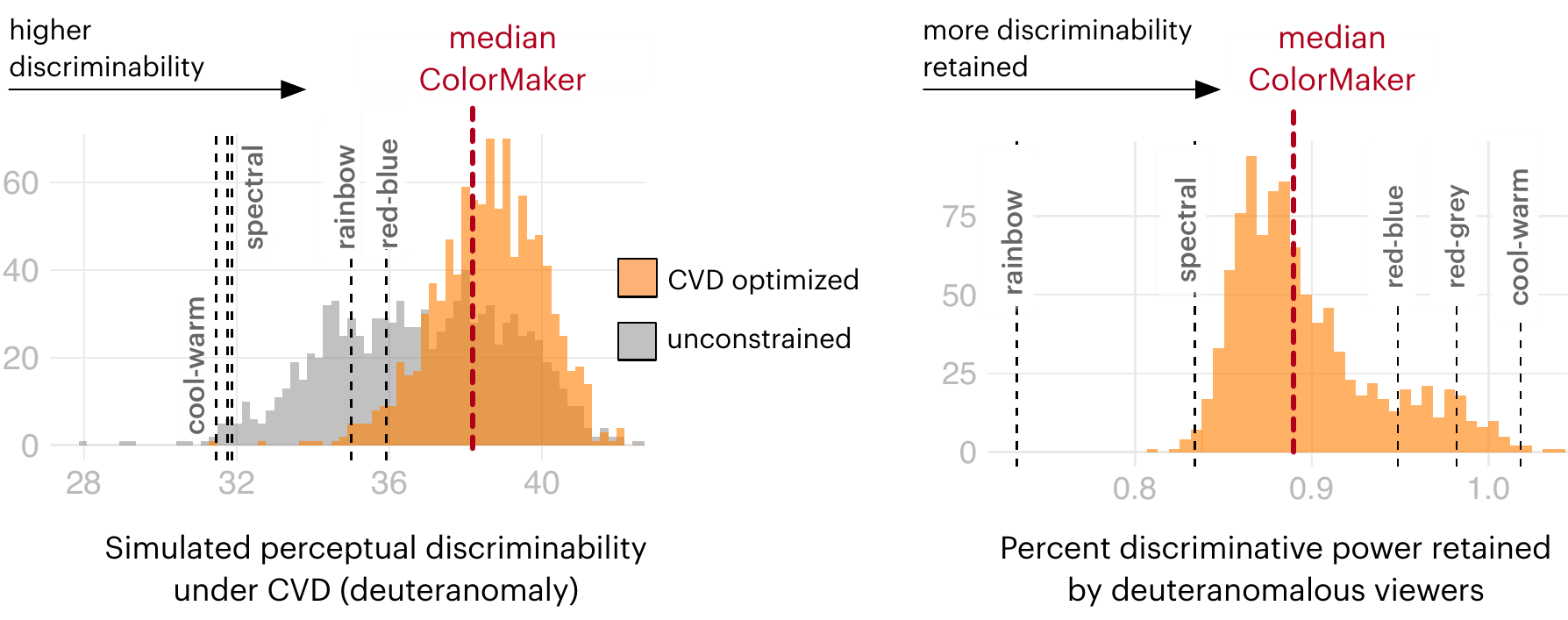}
    \caption{\emph{Left:} Simulated perceptual discriminability for a viewer with deuteranomaly. Colormaps were generated with either CVD-optimization enabled (yellow) or disabled (grey). \emph{Right}: Loss of discriminability due to CVD. A value of 1 indicates no theoretical loss.}
    \label{fig:cvd_discriminability}
    \Description[Difference in perceptual discriminability for CVD optimization]{
        Two histograms are shown, highlighting the difference in perceptual discriminability for CVD optimization. In the left histogram, we observe the results of discriminability under CVD optimization. An orange histogram represents the optimized results, whereas a gray histogram represents the unconstrained results. The optimized histogram is left-skewed. All the benchmark colormap indicators are positioned on the right side of the peak, indicating a higher level of discriminability as we progress to the right in the graph. In the second histogram, we take a look at the loss in discriminative power. Here, the histogram features a peak at approximately 0.9, indicating a loss of 88.6\% of their original color. Rainbow and Spectral colormaps are situated to the left of the peak, whereas the indicators for the other colormaps are to the right of the peak. Moving towards the right in the graph implies that more discriminability is retained.
    }
\end{figure*}

\subsubsection{Results}

Figure~\ref{fig:perceptual_quant} displays the results, featuring histograms of 1,000 generated designs each, obtained under the default optimization setting of 5,500 iterations. \hl{Red histograms illustrate high colorfulness (greater hue variation), while blue represents lower colorfulness (fewer hues).} We include comparable benchmarks in each plot to contrast their perceptual characteristics against ColorMaker's. Benchmark scores were calculated using the same equations, sampling the scale at $n$ equidistant points ($n=25$ for sequential and $n=31$ for diverging, as in ColorMaker's algorithm).

\vspace{1em}
\noindent\textbf{Perceptual Uniformity:} For all colormap types (Figure~\ref{fig:perceptual_quant}-A \& B), {perceptual uniformity} scores for ColorMaker are noticeably better than the benchmarks (lower score is better, as it implies smaller deviation from expected color distance). This is possibly due to the benchmarks having been designed using different distance metrics. For instance, \emph{blues} and \emph{red-grey} were made with equal Euclidean CIE Lab steps, whereas \emph{cool-warm} was created using a polar transformation of the CIE Lab space. That said, the narrow distributions suggest the algorithm can effectively optimize and maintain high perceptual uniformity across a variety of design styles.

\vspace{1em}

\noindent\textbf{Smoothness:}
As expected, the smoothness scores exhibit variations based on ColorMaker's generation parameters. For instance, in sequential colormaps, restricting hue variation results in distributions that largely overlap with the benchmark colormaps (Figure~\ref{fig:perceptual_quant}-C, blue). This is particularly true for colormaps like \emph{plasma} and \emph{viridis}. Interestingly, \emph{blues} appear smoother than the majority of ColorMaker's designs, possibly due to its straightforward, single-hue design. As the constraint on hue variation is relaxed, the generated scales tend to be less smooth (red histogram). This trend is mirrored in diverging colormaps (Figure~\ref{fig:perceptual_quant}-D), with the red distribution indicating lower smoothness compared to the benchmark colormaps. This suggests that, when hue variation is left unconstrained, the algorithm tends to favor more intricate and vibrant colormaps.

It is worth noting that, when the smoothness penalty is set to nearly maximum, we observe a long-tailed distribution in diverging scales (Figure~\ref{fig:perceptual_quant}-D, blue histogram). This might reflect occasional difficulty by the algorithm in maintaining fewer hues when indicated by the \hl{colorfulness} slider. However, this issue is relatively limited: out of 1,000 designs optimized for restricted hues, only 121 (12.1\%) scored 0.05 or worse in smoothness (for reference, Brewer's \emph{spectral}\cite{brewer1997spectral} scores approximately 0.04). Furthermore, ColorMaker offers an alternative mechanism for constraining hues through the preference shelf \hl{(see Figure} \ref{fig:usage_scanario1} \hl{for an example)}. Nevertheless, the majority of generated colormaps generally fall within the benchmarks for smoothness, with perhaps the exception of Moreland's \emph{cool-warm} -- the latter has been designed to be extraordinarily smooth~\cite{moreland2009diverging}, by way of softening the sharp, midpoint transition that is characteristic of many diverging colormaps (see Figure~\ref{fig:benchmarks}).

The results suggest that ColorMaker can generate a diverse range of single-hue, multi-hue, and diverging colormaps. Depending on the user-controlled \hl{colorfulness} parameter, the generated colormaps can either exhibit more vibrant colors or attain similar smoothness to comparable benchmarks. Notably, the algorithm appears capable of maintaining good perceptual uniformity even when set to favor high hue variation.

\vspace{1em}
\noindent\textbf{Perceptual Discriminability: } Sequential colormaps generated by ColorMaker demonstrate comparable discriminability to multi-hue benchmarks (Figure~\ref{fig:perceptual_quant}-E). In optimizations favoring hue variation (red), the distribution slightly shifts to the right, resulting in even higher discriminability. However, many of the generated colormaps still fall within the range of well-known multi-hue designs like \emph{viridis} and \emph{plasma}.

In contrast, the generated diverging scales consistently exhibit higher discriminability than the benchmarks (Figure~\ref{fig:perceptual_quant}-F). This suggests more vibrant designs, at times approaching the characteristics of rainbow scales. One potential reason for this difference is that traditional diverging scales often follow a conservative design, commonly concatenating two single-hued ramps ~\cite{wijffelaars2008generating}. Although the user can replicate this design by choosing two hues from the preference shelf and stretching them to overlap the two luminance arms, ColorMaker appears to generate more diverse and colorful designs when unconstrained by this strategy, resulting in higher discriminative power.

\vspace{1em}
\noindent\textbf{Discriminability for CVD Viewers: }Lastly, we examined the perceptual discriminability of diverging colormaps under CVD simulation. To conduct this analysis, we utilized equation~\ref{eq:discriminability}, replacing every color pair with their CVD-simulated appearance, as computed using the transform detailed in \S\ref{sec:cvd}. Figure~\ref{fig:cvd_discriminability}-left plots perceptual discriminability as experienced by a viewer with deuteranomaly. Compared to unconstrained generation, CVD optimization (yellow) shifts the distribution to the right, enhancing the accessibility of the generated colormaps. Notably, for the majority of optimized colormaps, the discriminative power as perceived by individuals with CVD surpasses that of comparable benchmarks like Brewer's \emph{red-blue}, known for its CVD-friendliness~\cite{harrower2003colorbrewer}.

\begin{figure*}[t]
    \centering
    \includegraphics[width=.7\textwidth]{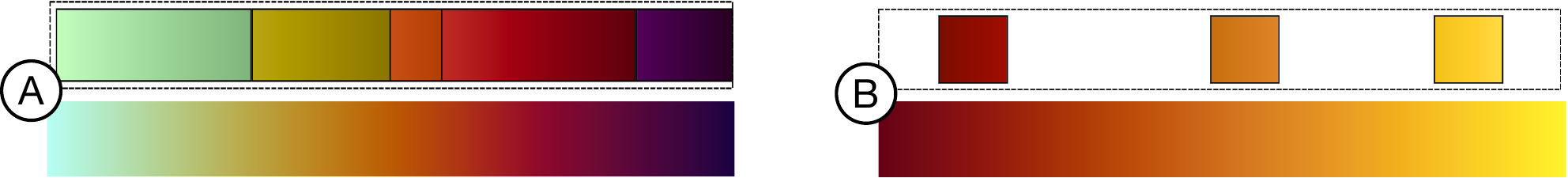}
    \caption{Participants employed two broad approaches when communicating their color preference: (A) `packing' the preference shelf to dictate precise expectations, or (B)  making partial specification and allowing the algorithm to populate the scale.}
    \label{fig:participant_strategy}
    \Description[Two varying methods used by participants for communicating color preference]{
        Two colormaps, each accompanied by a preference shelf above it are displayed. The first colormap is sequential, progressing through color from light green to yellow, and then to orange and finally to dark red. The preference shelf above it features five tightly packed color blocks: green, yellow, orange, red, purple. These blocks are positioned closely together without any gaps in between them. The second colormap is also sequential, transitioning from dark red to orange and then to yellow. The preference shelf above it has three color blocks: red, orange, and yellow. These blocks are placed apart from each other, with a lot of space within the shelf.
    }
\end{figure*}

Another approach to evaluating the impact of CVD on colormaps is to assess the theoretical loss in discriminative power (Figure~\ref{fig:cvd_discriminability}-right). Typical colormaps generated by ColorMaker retain approximately $88.6\%$ of their original color discriminability, which is higher than problematic colormaps like \emph{rainbow} ($73.3\%$) and \emph{spectral} ($83.4\%$), but lower than hand-crafted designs such as Brewer's \emph{red-blue} and \emph{red-grey} ($94.9\%$ and $98.2\%$ respectively), and \emph{cool-warm}, which experiences virtually no loss ($101\%$). This gap reflects a difference in strategy between ColorMaker and other tools, such as ColorBrewer, when it comes to ensuring CVD accessibility. While ColorBrewer follows a heuristic of avoiding large portions of the color space at a time (e.g., green tones~\cite{brewer1997spectral}), ColorMaker's optimization seeks a maximally accessible middle-ground solution. This results in more colorful designs that, although lose a larger percentage of their discriminative power, still manage to provide higher discriminability for CVD viewers.

\subsubsection{Summary}
Our analysis indicates that ColorMaker generates accessible designs with perceptual characteristics similar to known benchmarks. In some cases, the generated colormaps even surpass existing designs, achieving better uniformity and higher discriminative power. The range of scores, particularly for smoothness and discriminability, suggests that the algorithm can produce a diverse array of solutions. However, these results are for non-interactive generation, hence we evaluate the latter in a user study.

\subsection{User Study}

To evaluate ColorMaker's usability and expressive power, we conducted a user study with visualization practitioners and researchers who have varying levels of expertise in color design. The study included a set of structured and open-ended tasks, which were meant to simulate various color design scenarios.

\subsubsection{Participants}
 
We recruited 13 participants by email and by posting announcements in relevant forums, including slack channels frequented by computational scientists and visualization practitioners. We limited enrollment to those who create and/or consume visualizations regularly as part of their work. Participants were also required to have used at least one color-design tool in the past (e.g., ColorBrewer). Of the 13 participants we recruited, seven were student researchers in visualization, distributed systems, and neuroscience, four were visualization practitioners, and two were academic visualization researchers. Participants were compensated with a \$50 gift card. The study was approved by the Indiana University IRB.

\subsubsection{Procedures} 

The study was conducted remotely via Zoom. Participants interacted with ColorMaker through a standard web browser and shared their screens. We recorded screen content, video, and audio for analysis. We encouraged participants to think aloud and verbalize their design process. At the beginning of the study, we provided a brief tutorial to introduce the tool and demonstrate its various features. Participants then completed three color design tasks. In the first task, participants were asked to recreate two well-known colormaps (\emph{viridis}~\cite{viridis} and \emph{purple-green}~\cite{harrower2003colorbrewer}) in two separate trials. Reference color scales were displayed in a side panel. In the second task, participants were asked to design appropriate colormaps for three separate datasets that we provided: bathymetry data (scalar field), a slice from an MRI brain scan (scalar field), and air quality data (choropleth). Those datasets were shown on the side of the interface in lieu of the default example visualizations. The final task included a single open-ended trial, prompting participants to experiment and create a colormap they find pleasing, without specifying a target dataset. 
The study concluded with a semi-structured exit interview, in which participants described their general color design needs and commented on whether they thought ColorMaker would meet those.

\subsubsection{Results} 

Participants completed the study in 50.58 minutes on average ($\sigma=15.8$). Average trial-completion time was $6.09$ minutes in the first task ($\sigma = .9$). Tasks~2 and 3 took slightly less time, 5.03 ($\sigma = .48$) and $4.95$ ($\sigma = .87$) minutes on average, respectively. We analyzed the video and audio recordings to understand participants' usage patterns. We also analyzed their comments to understand what features participants were able to utilize effectively, and what challenges they faced while interacting with the tool. Several key themes emerged from the study:

\vspace{1em}\noindent\textbf{Colormap Generation and Ideation:} Several participants expressed satisfaction with the colormaps generated by the algorithm. For example, \emph{P13} remarked that ``the optimization under the hood does a great job of smoothing things out, especially when it gets to the final tweaks.'' Participants appreciated the tool's ability to ideate new designs, noting that the approach opens new design avenues for them. For instance, \emph{P9} said: ``Right now, the main way I tweak is to flip through the default [color] maps till I find something close and then adjust them by hand. I see myself using this [ColorMaker] to generate relatively quickly, entire new colormaps.'' Some remarked that the generated designs are immediately usable in their own visualizations: ``I'm pretty happy with the smoothness of the colors it generates. It will be good for heatmap-like data that I would use.'' (\emph{P3}).

\vspace{1em}\noindent\textbf{Specifying Preference:} Most participants found ColorMaker's model for specifying color preferences to be intuitive. This included the ability to add, move, `stretch', and remove colors. For example, \emph{P9} commented, ``The clicking, dragging, sliding [of color preferences] was all very intuitive. I enjoyed that.'' \emph{P5} also commented on the ease of eliminating unwanted colors from the scale: ``Some things are very easy to use, like if I want to trash this [color], I don't even have to worry about it and just drag it out and it's gone.''

There were broadly two different strategies used by participants to influence how the colormap should look. The first is to almost completely `pack' the preference shelf with all the colors the participant wanted to see (see Figure~\ref{fig:participant_strategy}-A). Four of the 13 participants employed this strategy of over-specifying the design. When asked about this approach, \emph{P8} said: ``I am thinking of the color in my head like when I am doing the visualization... If I have spaces, the algorithm may generate some different color which I am not imagining.'' This suggests a desire by some participants to exert fine-grained control over the algorithm. 

A second, more prevalent usage pattern was to provide an incomplete specification and let the algorithm ``fill in the gaps.'' Participants most commonly specified two or three desired colors and waited to see the initial optimization result (see Figure~\ref{fig:participant_strategy}-B for an example). Those who used this strategy generally thought the algorithm incorporated effective colors to complete the scale. For example, \emph{P1} remarked: ``After dragging and dropping a color, the generated color map makes sense and looks really pleasing, nice, smooth.'' Another participant, \emph{P2}, was pleasantly surprised with the result when, after adding only two colors to the preference shelf, was able to obtain a scale that matched their intuition: ``What's interesting is that it [the algorithm] introduced a new color that I was thinking of putting in there anyway.''

\vspace{1em}\noindent\textbf{Refining Generated Designs:} Participants used a variety of approaches to refine the generated designs. One frequent strategy was to adjust the preference shelf after an initial optimization run, by adding new colors or by repositioning and resizing existing color blocks. \emph{P13} commented on the ease of editing: ``I like the way the picker is set up in terms of being able to edit.... just like being able to tweak and tune.'' Different participants were also able to achieve similar design outcomes in different ways. For example, while \emph{P5} stretched a blue color to cover the entire scale (similar to the example in Figure~\ref{fig:usage_scanario1}), \emph{P7} achieved a similar colormap using two hues as input and making adjustments to the \hl{colorfulness} slider. In both cases, the algorithm responded by generating a single-hue, predominantly blue design. 

Several participants performed quick adjustments to the scale using the inline suggestion palette, which allows relative adjustment of hue or chroma at specific point in the scale. \emph{P13} commented on the usefulness of this feature: ``I really like the ability to get some suggestions when you're hovering [over the colormap] like this, because there were times where it's like, do I really want to tune out and pick a shade? And it's like no, I can just do this [hover] and get that suggestion. I thought that was really handy when in the drafting mode.''

\vspace{1em}\noindent\textbf{Optimization for CVD: } Of the 13 participants, seven used the built-in CVD optimization option to ensure accessibility. \emph{P10} commented on the usefulness of this feature to their visualization practice: ``I have to take into account if it is going to be printed and work well for CVD. So having the option when generating colors is nice...''. The participant went on to say: ``For the software I use, I don't have this [CVD] option, so it was good to have that. It was intuitive to use as well.'' Others also appreciated the ability to get an immediate CVD simulation: ``I really enjoyed the color vision checker because that's something I keep in mind and a lot of tools don't have anything like this.'' (\emph{P9}) and ``[the] CVD panel for this [ColorMaker] has more options. ColorBrewer has only one option: select colorblind safe maps and that's it.'' (\emph{P8}).

\vspace{1em}\noindent\textbf{Challenges and Suggestions for Improvement}: Despite the largely positive feedback, a few participants encountered challenges with the tool. One particular aspect of ColorMaker that seemed to confuse some is the luminance-as-hard-constraint model. This confusion often manifested during interaction with the preference shelf, and after a participant has added a color, only to see ColorMaker automatically adjust the luminance of that color to conform to the selected profile. Other participants had expected the tool to perform a linear interpolation between two colors by default. However, this was not typically the case, as the algorithm would frequently bend the colormap to improve smoothness while conforming to the user-specified hues. These examples seem to point to a possible gulf between some participants' mental models and the algorithm. The majority of participants appeared to grasp ColorMaker's mixed-initiative paradigm after a few interactions, and ultimately came to appreciate its expressiveness. However, the initial confusion experienced by some suggests more could be done to orient new users to the tool.

One feature that was used sparsely is the colormap curve editor (Figure~\ref{fig:ui}-E), which, for many, seemed too ``complex'' to ``mess with''. \emph{P2} said: ``I was a little bit intimidated by the curves. I have worked in curves before, but not all colors at once... Those were a bit tough for me to get into, at least at the first go, maybe [with] a bit more experience, I'd understand it better.'' This suggests a need to redesign this interaction to be more forgiving. 

Although ColorMaker was envisioned to fill a gap in design tools for continuous colormaps, several participants opined that they would have liked the inclusion of categorical and discrete, ordered palettes. Moreover, certain design features cannot be easily expressed with ColorMaker, which participants saw as a limitation. This includes color bands which were desired by some: ``I would like a sharper contrast between two colors almost like a contour itself, like a white line'' (\emph{P3}). Future work could add support for a wider variety of designs, including the ability to create sharp transitions along with support for rainbow-style colormaps (e.g., Turbo~\cite{turbo}).

One potential barrier to iterative refinement is the need to wait on the algorithm. On a laptop computer and with default generation parameters, it can take approximately eight seconds to complete a full optimization cycle (CVD modeling can take longer, though). To maintain interactivity, ColorMaker runs the optimization in the background and renders intermediate results, while also allowing the user to interrupt (e.g., with edits). One participant suggested a ``fast updating draft mode'' which would allow even faster design iteration. A current workaround involves decreasing the number of optimization iterations, which is controllable through a slider. Future work could focus on improving the efficiency of the algorithm, or on developing a parallelizable optimization and leveraging faster hardware (e.g., GPUs). 

\section{Limitations and Future Work}

Feedback from participants suggests that ColorMaker has successfully met the design requirements outlined in \S\ref{sec:design_requirements}. However, there are certain limitations to the current approach that warrant consideration. First, unlike techniques employing deterministic generation models~\cite{wijffelaars2008generating}, our algorithm relies on a stochastic, simulated annealing process. While the majority of generated colormaps appear to be comparable to benchmarks, there is a possibility of generating suboptimal designs. Therefore, it is essential for designers to evaluate the generated colormaps before adoption, including assessing the results of CVD simulation. The ColorMaker interface is designed to facilitate ideation, evaluation, and swift refinement of designs, but future enhancements could focus on ensuring more robust generation, possibly through automated post-optimization checks. These checks might involve comparing the generated colormap to known benchmarks (i.e., similar to the analysis in \S\ref{sec:analysis_perceptual}). Colormaps falling significantly short of benchmarks could be automatically discarded, with the algorithm generating new solutions. \hl{Similarly, although we offer the option to generate multiple colormaps simultaneously to promote ideation, the stochastic algorithm does not preclude the generation of repeated or highly similar designs. Future work could address this by measuring design similarity, and by introducing random biases to actively herd the algorithm into generating more diverse solutions.}

While ColorMaker strives to meet users' personal preferences, it currently lacks an \emph{empirical} model of color preference. In other words, the optimization does not consider whether a particular colormap is aesthetically pleasing. Aesthetics is essential, especially because designers often base their color choices on visual appeal~\cite{lau2007towards,cawthon2007effect}. Future work could incorporate an empirical aesthetics model into the objective scoring function. Moreover, there is room to optimize for additional cognitive factors, like color nameability~\cite{reda2021color, heer2012color}, thus enabling the generation of designs with more distinct and recognizable color names. 

\hl{Lastly, the ColorMaker algorithm currently depends on several parameters, including the number of control points ($n$). These parameters are operationalized into the objective cost functions (e.g., the smoothness penalty function uses $n$ and $\frac{n}{2}$ samples to estimate curvature), and will therefore influence the quality of the results. Future work should attempt to characterize the impact of these parameters on colormap generation and, where possible, reduce their number to a minimum.} The current implementation is also limited in the range of luminance profiles (and consequently colormap designs) it can generate. We believe that this limitation can be overcome with minor adaptations. Future implementations could also permit some flexibility in meeting the requested constraints, including minor deviations from a strict luminance profile. Such pliability would allow for the generation of a wider diversity of designs, including rainbows, which have merit in certain cases~\cite{reda2022rainbow,reda2019evaluating,ware2023rainbow}.

\section{Conclusion}

Although numerous tools cater to the creation of categorical color palettes, designing quantitative colormaps for visualization is less supported. This gap led us to introduce ColorMaker, a mixed-initiative approach for the generation and customization of continuous colormaps. ColorMaker employs simulated annealing to produce a diverse range of colormap styles, including CVD-friendly designs. It allows for incorporating user color preferences, enabling iterative colormap refinement. Evaluation confirms that this approach consistently delivers high-quality colormaps, on par with established designs. Furthermore, a user study demonstrates ColorMaker's intuitiveness and its potential for empowering visualization designers to create colormaps for various needs. 

\begin{acks}
This paper is based upon research supported by the National Science Foundation under award 1942429. MEP and KR were also supported in part by the Office of Science, U.S. Department of Energy, under contract DE-AC02-06CH11357. KL and YW are supported in part by NSF China (No. 62132017, 62141217), and the Shandong Provincial Natural Science Foundation (No. ZQ2022JQ32). 
\end{acks}

\interlinepenalty=10000
\balance
\bibliographystyle{ACM-Reference-Format}
\bibliography{color.bib}

\end{document}